\documentclass[fleqn,usenatbib]{mnras}

\usepackage{newtxtext,newtxmath}
\usepackage[T1]{fontenc}

\DeclareRobustCommand{\VAN}[3]{#2}
\let\VANthebibliography\thebibliography
\def\thebibliography{\DeclareRobustCommand{\VAN}[3]{##3}\VANthebibliography}

\usepackage{graphicx}	% Including figure files
\usepackage{amsmath}	% Advanced maths commands
\usepackage{enumitem}

\usepackage{soul}
\usepackage[dvipsnames]{xcolor}

\usepackage{ulem}

\newcommand{\removed}[1]{\relax}
\newcommand{\inserted}[1]{#1}

\newcommand{\erased}[1]{\relax}
\newcommand{\textnew}[1]{#1}

\newcommand{\newerased}[1]{\relax}
\newcommand{\newinserted}[1]{#1}

\usepackage{booktabs}

\title[Carbon stars as standard candles]{Carbon stars as standard candles - III. Un-binned maximum likelihood fitting and comparison with TRGB estimations }

\author[Parada et al.]{
Javiera Parada,$^{1}$\thanks{E-mail: jparada@phas.ubc.ca}
Jeremy Heyl,$^{1}$
Harvey Richer,$^{1}$
Paul Ripoche$^{1}$
\newauthor{ and Laurie Rousseau-Nepton$^{2,3}$}
\\
$^{1}$Department of Physics and Astronomy, University of British Columbia, 6224 Agricultural Road, Vancouver, British Columbia V6T 1Z4, Canada\\
$^{2}$ Canada-France-Hawaii Telescope, Kamuela, HI, 96743, USA\\
$^{3}$ Department of Physics and Astronomy, University of Hawaii at Hilo, Hilo, HI, 96720, USA
}

\date{Accepted XXX. Received YYY; in original form ZZZ}

\pubyear{2021}

\begin{document}
\label{firstpage}
\pagerange{\pageref{firstpage}--\pageref{lastpage}}
\maketitle

\begin{abstract}
In the second paper of this series, we developed a new distance determination method using the median $J$ magnitude of carbon-rich asymptotic giant branch stars (CS) as standard candles and the Magellanic Clouds as the fundamental calibrators. The $J$-band CS luminosity function was modeled using a modified Lorentzian distribution whose parameters were used to determined whether the LMC or SMC was the most suitable calibrator. In this third paper of the series, we expand our sample of galaxies and introduce a more robust method to determine the parameters of the Lorentzian model. The new fitting method uses an un-binned maximum likelihood estimator to determine the parameters of the Lorentzian model resulting in parameter errors that are significantly smaller compared to the second paper. We test our method in NGC 6822, IC 1613, NGC 3109 and WLM. We also estimate the distances to the same sample of galaxies via the tip of the red giant branch (TRGB) detection method. Our results from the CS measurements agree well with those obtained from the TRGB.

\end{abstract}

\begin{keywords}
stars: AGB -- stars: carbon -- galaxies: distances
\end{keywords}

%%%%%%%%%%%%%%%%%%%%%%%%%%%%%%%%%%%%%%%%%%%%%%%%%%

%%%%%%%%%%%%%%%%% BODY OF PAPER %%%%%%%%%%%%%%%%%%

\section{Introduction}
\label{sec:intro}

The main tension with the Hubble Constant, $H_0$, is the discrepancy between the values based on the standard cosmological model, with those obtained using observations of the local universe with differences up to $\sim6\sigma$ \citep{2019NatAs...3..891V, 2021ApJ...908L...6R}. While the latest results obtained on the cosmic scale agree with each other \citep{2020JCAP...12..047A, 2020A&A...641A...6P}, the values of $H_0$ determined on the short-distance scale have a tension of their own \citep{2021ApJ...919...16F}. This inconsistency has encouraged a large number of research groups to develop or improve different distance determination methods to increase the accuracy of the local measurement of $H_0$, e.g. the tip of the red giant branch (TRGB) \citep{2020ApJ...891...57F}, Cepheids \citep{2021ApJ...908L...6R}, Mira variables \citep{2020ApJ...889....5H}, surface brightness fluctuations \citep{2021A&A...647A..72K, 2021ApJ...911...65B}. 

% Local determinations of $H_0$ vary depending on the standard candle used to calibrate the extragalactic distance scale \comment{reference}

A potential standard candle that has recently attracted some attention is carbon rich asymptotic giant branch stars (AGB CS) \citep{paper1, 2020ApJ...899...66M, 2020ApJ...899...67F,paper2,2021ApJ...907..112L,2021ApJ...916...19Z}, or J-region stars as defined by \cite{2000ApJ...542..804N} and \citet{2001ApJ...548..712W}. CS are formed when AGB stars experience sufficient carbon dredge-up to make the carbon-to-oxygen ratio in the stellar atmosphere greater than unity. This excess carbon near the surface produces carbon-containing molecules making the star redder than the bulk of the AGB population. Their red colour and high luminosity make CS a distinctive and easily identifiable feature in the near-infrared (NIR) colour-magnitude diagram (CMD) (see Fig. \ref{fig:CSselMCs}).

CS were first used as a distance indicator by \cite{richer_1984}. By assuming that CS have the same mean luminosity in any galaxy, \cite{richer_1984} determined the distance to NGC 205 using the CS in the Magellanic clouds as calibrators. \cite{2020ApJ...899...66M}, \cite{2020ApJ...899...67F} and \cite{2021ApJ...907..112L} adopt a similar approach by defining an absolute mean $J$ magnitude for CS based on the absolute mean magnitudes of the CS in the Large and Small Magellanic clouds (LMC and SMC). \cite{2021ApJ...916...19Z} also use the CS mean $J$ magnitude to determine distances with only the LMC as the fundamental calibrator. In their method, they first fit the luminosity function with a custom profile composed of the superposition of a Gaussian and a quadratic function. The mean $J$ magnitude is then defined as the mean of the Gaussian component of the profile. In our series of papers we instead use the median $J$ magnitude of the CS and either the LMC \textit{or} SMC as calibrators, determined on a case by case basis.

In Paper I \citep{paper1} we showed how the CS region ($J-K$)$_0$ colour limits were determined, and analysed the luminosity function of the colour-selected CS. A significant difference was found between the median $J$ magnitude of the CS in the LMC and the SMC, attributed to the difference in metallicity between the two Clouds \citep[see section 3.4 of][and references therein]{paper2}. In Paper II \citep{paper2}, we developed a distance determination method to Magellanic type galaxies based on the median $J$ magnitude of the CS and tested it in two Local Group galaxies. To compensate for metallicity effects, the distance is determined using the LMC \textit{or} SMC as fundamental calibrator, depending on the characteristics of the luminosity function of the CS of the target galaxy, specifically on how skewed the luminosity function is (see section 4.1 of Paper II). To determine this skewness, we fitted the binned CS luminosity functions with a modified Lorentzian distribution. 

In this paper (the third of the series) we extend our sample of galaxies and develop an un-binned maximum likelihood fitting method to determined the parameters of the Lorentzian distribution. Additionally, we estimate the distances to our target galaxies using the TRGB method and do a comparison with the results obtained from the CS. Comparing results obtained with two different methods on the same data sets will allow us to test the precision and accuracy of the CS method and compare its performance against the TRGB.

%We will use the distance to the LMC determined by \citet{2019Natur.567..200P} based on 20 eclipsing binary systems that is accurate to $1 \%$. They found a mean distance modulus: $\mu_\text{LMC} = 18.477\pm 0.004$ (statistical) $\pm~0.026$ (systematic).

%%%%%%%%%%%%%%%%%%%%%%%%%%%%%%%%%%%%%%%%%%%%%%%%%%%%%%%%%%%%%%%%
%%%%%%%%%%%%%%%%%%%%%%%%%%%%%%%%%%%%%%%%%%%%%%%%%%%%%%%%%%%%%%%%
%%%%%%%%%%%%%%%%%%%%%%%%%%%%%%%%%%%%%%%%%%%%%%%%%%%%%%%%%%%%%%%%
% 
%
%  _____        _                               _           _     
% |  __ \      | |            /\               | |         (_)    
% | |  | | __ _| |_ __ _     /  \   _ __   __ _| |_   _ ___ _ ___ 
% | |  | |/ _` | __/ _` |   / /\ \ | '_ \ / _` | | | | / __| / __|
% | |__| | (_| | || (_| |  / ____ \| | | | (_| | | |_| \__ \ \__ \
% |_____/ \__,_|\__\__,_| /_/    \_\_| |_|\__,_|_|\__, |___/_|___/
%                                                  __/ |          
%                                                 |___/           
%
\section{Data Analysis \& Catalogues}
\label{sec:data}

%%%%%%%%%%%%%%%%%%%%%%%%%%%%%%%%%%%%%%%%%%%%%%%%%%%%%%%%%%%%%%
\subsection{Magellanic Clouds}
\label{subsec:mcdata}

Based on data obtained from the Two Micron All Sky Survey (2MASS) \footnote{\url{https://irsa.ipac.caltech.edu/Missions/2mass.html}} \citep{2006AJ....131.1163S}, we constructed individual catalogues for the LMC and SMC. 2MASS photometric data are available in three near-infrared bands; $J$ (1.25 microns), $H$ (1.65 microns), and $K_{s}$ (2.17 microns). Galactic foreground stars were removed by cross-matching the 2MASS stars with those found in \textit{Gaia} EDR3 \citep{GaiaMission, GaiaEDR3}. True members of the LMC and SMC were then selected using Gaia EDR3 proper motion measurements. A detailed description of this process can be found in sections 2.1 and 3.1 of Paper I. 

The Magellanic Clouds will serve as our fundamental distance calibrators, therefore we transform the 2MASS apparent magnitudes to absolute magnitudes. For the LMC, we adopt the mean distance modulus determined by \citet{2019Natur.567..200P}: $\mu_{\mathrm{LMC}}$ = 18.477 $\pm$ 0.004 (statistical) $\pm$ 0.026 (systematic). For the SMC, we take the distance modulus found by \citet{2020ApJ...904...13G}: $\mu_{\mathrm{SMC}}$ = 18.977 $\pm$ 0.016 (statistical) $\pm$ 0.028 (systematic). In Papers I and II the distance modulus for the SMC was taken from \citet{2016ApJ...816...49S} ($\mu_{\mathrm{SMC}}$ = 18.96 $\pm$ 0.03 (stat) $\pm$ 0.05 (sys)). In this paper we will update our previous results to reflect the new SMC distance modulus value. \removed{We finally correct the magnitudes for reddening and extinction. For this, we use the reddening maps for both the Magellanic Clouds obtained by \citet{2020ApJ...889..179G} and the extinction coefficients for the 2MASS bands from \citet{2003ApJ...594..279G}.}

\begin{figure}
\centering
\includegraphics[width=\columnwidth]{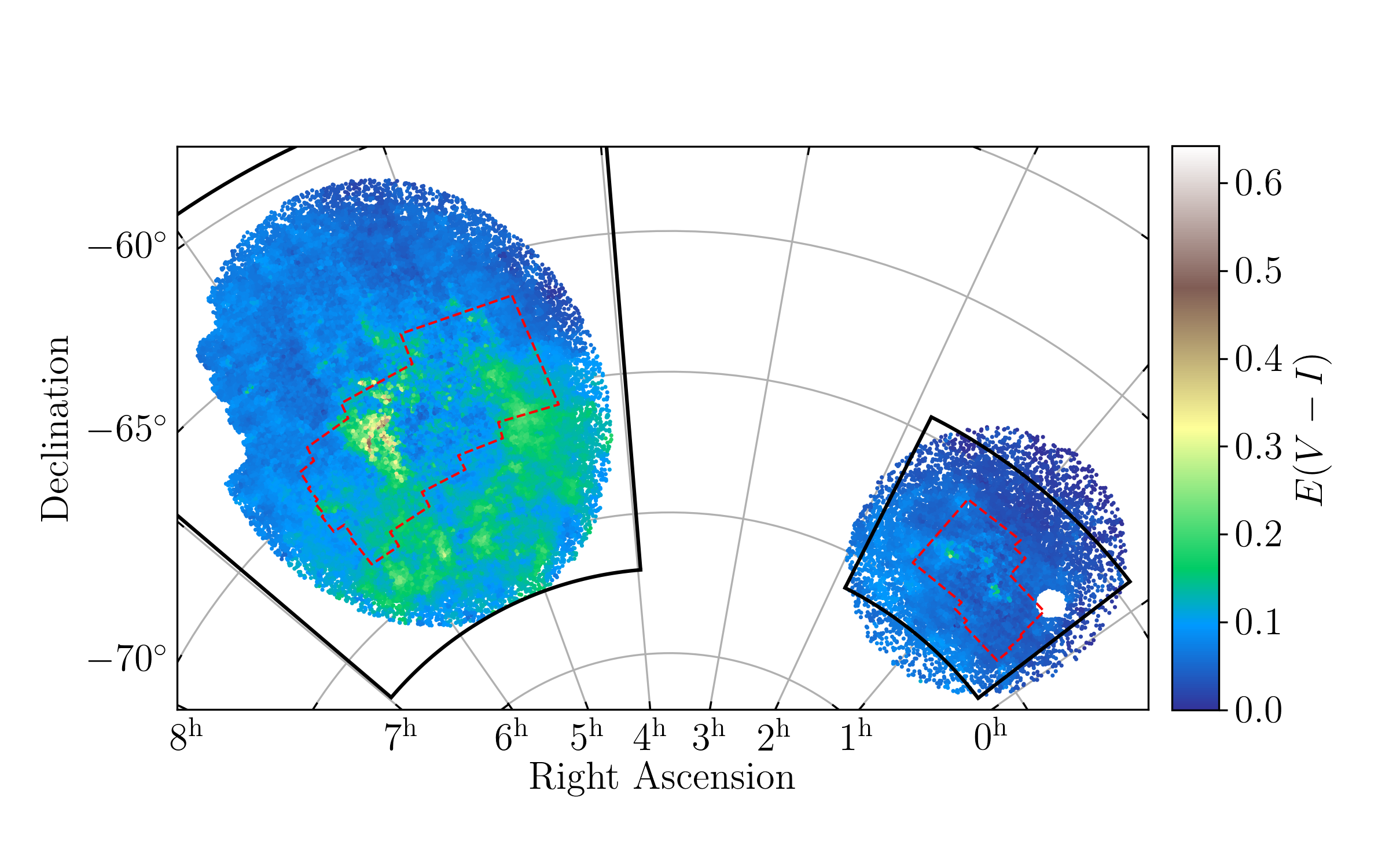}
\caption{Reddening map for the Magellanic Clouds. The colour excess $E(V-I)$ for each star is retrieved from the \citet{2021ApJS..252...23S} reddening map. The black lines delimit the original fields used in Papers I and II. The dashed red lines show the boundaries of the reddening maps from \citet{2020ApJ...889..179G} which are considerably smaller than the original fields.}
\label{fig:redmaps}
\end{figure}

\inserted{Finally, we correct the magnitudes for reddening and extinction. In the previous papers of this series the Magellanic Clouds data were corrected using the mean $E(B-V)$ reddening values from the \citet{2020ApJ...889..179G} reddening maps along with \citet{2003ApJ...594..279G} extinction factors. Since then, new reddening maps and extinction factors for the Magellanic Clouds have been published. \citet{2021ApJS..252...23S} developed new reddening maps\footnote{Publicly available at \url{https://ogle.astrouw.edu.pl/cgi-ogle/get_ms_ext.py}} that cover a significantly larger area compared to \citeauthor{2020ApJ...889..179G}'s maps. As shown in Fig. \ref{fig:redmaps}, for the SMC, the extent of \citeauthor{2021ApJS..252...23S}'s map covers nearly the entire original SMC field. In the case of the LMC, \citeauthor{2021ApJS..252...23S}'s map covers a slightly smaller area compared to the original data, nevertheless over 95\% of the CS selected in Papers I and II fall within the limits of \citeauthor{2021ApJS..252...23S}'s map. The extent of the new maps allows us to obtain a location dependent reddening value for each individual star without significant loss of stars.}

\inserted{In Fig. \ref{fig:redmaps} \citeauthor{2020ApJ...889..179G}'s maps are centred in the areas of the LMC and SMC where the reddening is higher compared to other parts of the galaxy. The mean reddening values found from both maps are presented in Table \ref{tab:mean_red}, \citeauthor{2021ApJS..252...23S}'s maps show a much lower mean but this number is driven by the fact that the maps include the outskirts of the galaxies where reddening is very low. In \citeauthor{2021ApJS..252...23S} the authors compared their maps to  \citeauthor{2020ApJ...889..179G}'s maps using the intrinsic colour of red clump stars found in the two studies. In the overlapping areas \citeauthor{2021ApJS..252...23S} find red clump stars to have on average redder intrinsic colours compared to those found by \citeauthor{2020ApJ...889..179G},  with differences ($(V-I)_{0,\mathrm{Skowron}} - (V-I)_{0,\mathrm{Gorski}}$) of 0.045 and 0.048 magnitudes in the LMC and SMC respectively. If the stars are found to be intrinsically redder, then the reddening correction needed to be applied to the observed data is smaller.}

\begin{table}
    \centering
        \caption{Mean reddening values for the LMC and SMC from the \citet{2020ApJ...889..179G} (G20) and \citet{2021ApJS..252...23S} (S21) reddening maps. A lower mean value from S21 is expected given their map covers also the outskirts of the galaxies where reddening is very low. The expression used in the fourth column is taken from S21.}
    \begin{tabular}{c c c c}
        \midrule
             & G20 & S21 & S21 \\
             & $E(B-V)$ & $E(V-I)$ & $E(B-V) = \frac{(V-I)}{1.318}$ \\
        \midrule
        \midrule
         LMC & 0.127    &  0.100   &  0.076   \\
         SMC & 0.084    &  0.047   &  0.036   \\
        \midrule
    \end{tabular}
    \label{tab:mean_red}
\end{table}

\inserted{The extinction coefficients ($R_x$) are also updated with the values recently published by \citet{2022MNRAS.511.1317C}. The $R_x$ values from \citet{2003ApJ...594..279G} and  \citeauthor{2022MNRAS.511.1317C} are compared in Table \ref{tab:ext_factors}. The extinction coefficients for $J$ and $K$ are higher in \citeauthor{2022MNRAS.511.1317C} for both the LMC and SMC. Most importantly, the ratio between $R_J$ and $R_K$ changes drastically with $R_J/R_K\sim8$ for \citeauthor{2003ApJ...594..279G} and only $R_J/R_K\sim2.5$ for \citeauthor{2022MNRAS.511.1317C}. This will have a impact on the analysis, especially on the colour dependent magnitudes used to obtained the TRGB magnitude. }

\begin{table}
    \centering
        \caption[Extinction coefficients for the LMC and SMC]{Extinction coefficients for the LMC and SMC from \citet{2003ApJ...594..279G} and \citet{2022MNRAS.511.1317C}.}
    \begin{tabular}{c c c c c c c}
        \midrule
             & \multicolumn{3}{c}{\citet{2003ApJ...594..279G}} & \multicolumn{3}{c}{\citet{2022MNRAS.511.1317C}} \\
             & $R_J$ & $R_H$ & $R_K$ & $R_J$ & $R_H$ & $R_{K_s}$ \\
        \midrule
        \midrule
         LMC & 0.877 & 0.635 & 0.102 & 1.015 & 0.640 & 0.413   \\
         SMC & 0.378 & 0.487 & 0.046 & 0.742 & 0.468 & 0.302  \\
        \midrule
    \end{tabular}
    \label{tab:ext_factors}
\end{table}

\begin{figure}
\centering
\includegraphics[width=\columnwidth]{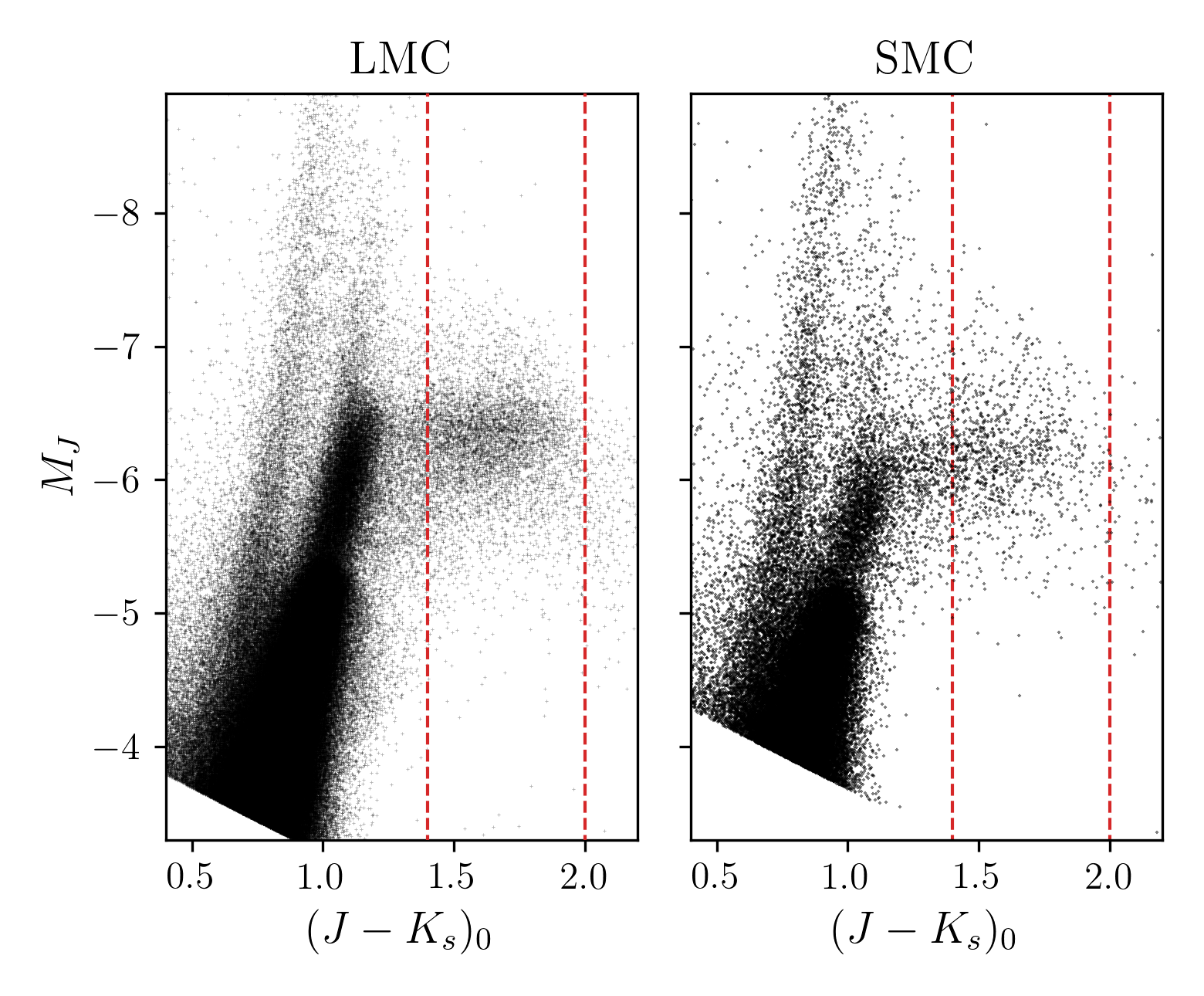}
\caption{$(J-K_{s}$)$_{0}$, M$_{J}$ colour-magnitude diagrams for the Large (\textit{left}) and Small (\textit{right}) Magellanic Clouds. The data shown are the resulting catalogues after filtering foreground contamination and correcting for reddening. The vertical red lines represent the colour limits for the carbon star region (1.4$<$ $(J-K_{s})_{0}$ $<$ 2.0).}
\label{fig:CSselMCs}
\end{figure}

The resulting foreground cleaned $(J-K_{s})_{0}$, M$_{J}$ CMD for the LMC and SMC are shown in Fig. \ref{fig:CSselMCs}.

%%%%%%%%%%%%%%%%%%%%%%%%%%%%%%%%%%%%%%%%%%%%%%%%%%%%%%%%%%%%%%%%
\subsection{NGC 6822, IC 1613 \& WLM (WIRCam data)}
\label{sec:wircamdata}

NGC 6822 and IC 1613 were the target galaxies in Paper II. The data for IC 1613 used in Paper II come from \cite{2015A&A...573A..84S} and are publicly available through VizieR\footnote{\url{http://cdsarc.unistra.fr/viz-bin/cat/J/A+A/573/A84}}. For the current paper the data for NGC 6822 remain the same as Paper II while for IC 1613 we introduce a new data set obtained as explained below. We will compare the results from this new data set with the ones from Paper II. 

The NGC 6822, IC 1613 and WLM data come from observations made with the Wide-field InfraRed Camera (WIRCam) at the Canada-France-Hawaii Telescope (CFHT) using three different filters: $J$ (1.253 $\mu$m), $H$ (1.631 $\mu$m), and $K_{s}$ (2.146 $\mu$m). The individual exposure times in each filter are 60, 15 and 25 seconds in $J$, $H$ and $K_s$ respectively. The number of images and total exposure times for each galaxy are summarized in Table \ref{tab:wircamdata}.
 
For each galaxy and filter we created a co-added image on which point spread function (PSF) photometry was performed using DAOPHOT II \citep{daophot} and ALLSTAR \citep{allstar}. Details of the data reduction and photometry process are presented in section 2.2.1 of Paper II. For each galaxy the individual $J$, $H$ and $K_s$ catalogues were combined into a single catalogue where only the stars found in all three filters remained.

The calibrators' (the LMC and SMC) photometry comes from 2MASS. The $JHK_{s}$ filters used by 2MASS are not identical to the WIRCam $JHK_{s}$ filters. To transform the WIRCam photometry into the 2MASS system we use the colour term equations provided by the WIRwolf image stacking pipeline\footnote{\url{http://www.cadc-ccda.hia-iha.nrc-cnrc.gc.ca/en/wirwolf/docs/filt.html}}:
\begin{equation}
\begin{split}
&J_{\mathrm{2MASS}} = J_{\mathrm{WIRCam}} + 0.071 \times (J-H)_{\mathrm{WIRCam}} \\
&H_{\mathrm{2MASS}} = H_{\mathrm{WIRCam}} - 0.034 \times (J-H)_{\mathrm{WIRCam}} \\
&K_{\mathrm{2MASS}} = K_{\mathrm{WIRCam}} - 0.062 \times (H-K)_{\mathrm{WIRCam}}.\\
& \quad \quad \qquad +0.002 \times (J-H)_{\mathrm{WIRCam}}.
\end{split}
\end{equation}

After accounting for the difference in the $JHK_{s}$ filter systems, we calibrate the instrumental magnitudes by taking the mean magnitude difference between stars found in both 2MASS and in the fields observed for our galaxies. Additional details on the calibration can be found in section 2.2.1 of Paper II.

The resulting CMDs show a considerable amount of contamination from background galaxies, particularly noticeable in the colour range defined for the CS. Background galaxies reveal themselves as extended sources compared to stars. ALLSTAR provides two indices to help identify non-stellar sources, a sharpness (SHARP) and a goodness-of-fit (CHI) parameter. Using these we remove the sources that are most likely non-stellar. As detailed in section 2.2.2 of Paper II, we use red giant stars to obtain a mean reference value of SHARP and CHI and consider any source within $\pm2.5\sigma$ of the mean value to be a well-defined point-like source.

The final correction to the data is for reddening. NGC 6822 is located at a galactic latitude of about $-18^\circ$, which results in a considerable amount of reddening. We adopt $E(B-V)=0.35$ defined by \citet{2014ApJ...794..107R} which includes both the foreground reddening and an estimate of the internal reddening within NGC 6822. For IC 1613 and WLM we use the average line-of-sight extinction estimated by \cite{2011ApJ...737..103S}: $E(B-V)=0.021$ and $E(B-V)=0.03$ respectively.

The final calibrated and cleaned CMDs are shown in Fig. \ref{fig:CSsel}.

% TABLE: WIRCAM OBSERVATIONS
\begin{table}
\caption{WIRCam near-infrared observational details. Individual frames exposure times are 60, 15 and 25 seconds for $J$, $H$ and $K_s$ respectively.}
\centering
\begin{tabular}{l c c c c c c}
\hline
Galaxy & \multicolumn{3}{c}{N$^o$ of exposures} & \multicolumn{3}{c}{Total exp. t (sec)} \\
& $J$ & $H$ & $K_s$ & $J$ & $H$ & $K_s$ \\
\hline
\hline
NGC 6822 & 40 & 160 & 120 & 2400 & 3000 & 3000 \\
IC 1613  & 20 & 84 & 60 & 1260 & 1260 & 1500 \\
WLM      & 19 & 80 & 60 & 1200 & 1140 & 1500 \\
\hline
\end{tabular}
\label{tab:wircamdata}
\end{table}

%%%%%%%%%%%%%%%%%%%%%%%%%%%%%%%%%%%%%%%%%%%%%%%%%%%%%%%%%%%%%%%%
\subsection{IC 1613 (Sibbons et al. data)}
\label{sec:ic1613sibdata}

Our understanding of the IC 1613 data published by \cite{2015A&A...573A..84S} in Paper II, was incorrect. The data available through VizieR\footnote{\url{http://cdsarc.unistra.fr/viz-bin/cat/J/A+A/573/A84}}, came from observations made with the Wide Field CAMera (WFCAM) mounted on UKIRT. The reduction and calibration process of \citeauthor{2015A&A...573A..84S} uses 2MASS data  but the magnitudes presented in the catalogue are in the WFCAM system (not 2MASS as stated in Paper II). The data in the catalogue has also already been corrected for foreground reddening using the \cite{1998ApJ...500..525S} extinction maps (because we missed this detail in Paper II, we applied the correction again ending with an overly de-redden data set). \citeauthor{2015A&A...573A..84S} gives a range for the reddening of $E(B-V)=0.02-0.03$ but does not specify the exact value applied to the catalogue. Assuming an extinction to reddening ratio in the V-band of 3.1, the difference between the minimum and maximum reddening used in \citeauthor{2015A&A...573A..84S} would translate to a difference in the extinction in $J$ of $\Delta A_J\approx0.01$ magnitudes. 

To transform \citeauthor{2015A&A...573A..84S} data from the UKIRT to the 2MASS system we utilize the transformation equations given by \cite{2009MNRAS.394..675H}:

\begin{equation}
\begin{split}
&J_{\mathrm{2MASS}} = J_{\mathrm{WFCAM}} + 0.075 \times (J-H)_{\mathrm{WFCAM}} - 0.002 \\
&H_{\mathrm{2MASS}} = H_{\mathrm{WFCAM}} - 0.081 \times (J-H)_{\mathrm{WFCAM}} + 0.032 \\
&K_{\mathrm{2MASS}} = K_{\mathrm{WFCAM}} - 0.010 \times (J-K)_{\mathrm{WFCAM}}.\\
& \quad \quad \qquad +0.001 \times (J-H)_{\mathrm{WFCAM}}.
\end{split}
\end{equation}

%%%%%%%%%%%%%%%%%%%%%%%%%%%%%%%%%%%%%%%%%%%%%%%%%%%%%%%%%%%%%%%%
\subsection{NGC 3109}

The data for NGC 3109 come from observations obtained by  \cite{2006ApJ...648..375S}. These observations were made with the Infrared Spectrometer And Array Camera (ISAAC; decommissioned in 2013) mounted on the 8.2m ESO Very Large Telescope (VLT). The images were downloaded from the ESO Science Archive Facility and processed with Jitter from the Eclipse package \citep{1997Msngr..87...19D} developed by ESO for the reduction of near-infrared data. The observations consist of three fields (F1, F2 and F3). Each field was observed on two different nights, except for F1 which was observed only once. The images were obtained using just two near-infrared filters: $J$ (1.25 microns) and $K_{s}$ (2.16 microns). Some of the observed frames have large spatial offsets from the reference frames and/or are unfocused. When co-adding the images we chose to remove these frames, therefore the total exposure time for each field varies. Each frame has an exposure time of 30 and 15 seconds in $J$ and $K_{s}$ respectively. The total number of frames and exposure time used in this paper for each field are detailed in table \ref{tab:isaacdata}

% TABLE: ISAAC OBSERVATIONS
\begin{table}
\caption{ISAAC near-infrared observations details for each field observed in NGC 3109. Individual frame exposure times are 30 and 15 seconds for $J$ and $K_{s}$ respectively.}
\centering
\begin{tabular}{c c c c c}
\hline
Field & \multicolumn{2}{c}{N$^o$ of exposures} & \multicolumn{2}{c}{Total exp. t (sec)} \\
& $J$ & $K_s$ & $J$ & $K_s$ \\
\hline
\hline
F1 & 12 & 38 & 360 & 570 \\
F2 & 17 & 65 & 510 & 975 \\
F3 & 19 & 68 & 570 & 1020\\
\hline
\end{tabular}
\label{tab:isaacdata}
\end{table}

PSF photometry was performed independently in each co-added image using the same method as for the WIRCam data. For each field we construct a photometry file which contains only the stars found in both the $J$ and $K_{s}$ images. These catalogues were then cleaned using the SHARP and CHI parameters output by DAOPHOT II as specified in section \ref{sec:wircamdata}. 

Before calibrating the NGC 3109 photometry, it is necessary to bring all three fields to a common instrumental zero-point. Both F1 and F3 partially overlap with F2 allowing us to find common stars between F1 and F2, and F3 and F2. For each common star the magnitude difference is calculated. The mean of the magnitude difference is then used to adjust F1 and F3 zero-points relative to F2.

The $JHK_{s}$ filters used by ISAAC closely match the $JHK_{s}$ filters defined in the Las Campanas Observatory (LCO) (or Persson standards; \citep{1998AJ....116.2475P}) system. Therefore colour terms between the two systems are negligible \citep{isaaclco, 2011A&A...528A..43L}. Comparison between the LCO and 2MASS systems show that the colour terms do not differ from zero by more than a few thousandths of a magnitude \citep{2001AJ....121.2851C}. For our purposes we consider the colour terms between ISAAC and 2MASS $JHK_{s}$ system negligible and do not apply any transformation. 

To calibrate the NGC 3109 data, we follow the same procedure as for the WIRCam data sets (see Sec. \ref{sec:wircamdata}). Stars from 2MASS are matched to stars found within the three observed fields, and the weighted mean of the magnitude differences is added to the ISAAC data. Once the data has been zero-pointed to 2MASS, we add the reddening correction: $E(B-V)=0.056$ \citep{2011ApJ...737..103S}. The final CMD is shown in Fig. \ref{fig:CSsel}.

%%%%%%%%%%%%%%%%%%%%%%%%%%%%%%%%%%%%%%%%%%%%%%%%%%%%%%%%%%%%%%%%
%%%%%%%%%%%%%%%%%%%%%%%%%%%%%%%%%%%%%%%%%%%%%%%%%%%%%%%%%%%%%%%%
% ╭━━━╮       ╭╮           ╭━━━╮ ╭╮
% ┃╭━╮┃       ┃┃           ┃╭━━╯╭╯╰╮
% ┃┃ ╰╯╭━━╮╭━╮┃╰━╮╭━━╮╭━╮  ┃╰━━╮╰╮╭╯╭━━╮╭━╮╭━━╮
% ┃┃ ╭╮┃╭╮┃┃╭╯┃╭╮┃┃╭╮┃┃╭╰╮ ╰━━╮┃ ┃┃ ┃╭╮┃┃╭╯┃ ━┫
% ┃╰━╯┃┃╭╮┃┃┃ ┃╰╯┃┃╰╯┃┃┃┃┃ ╭━━╯┃ ┃╰╮┃╭╮┃┃┃ ┣━ ┃
% ╰━━━╯╰╯╰╯╰╯ ╰━━╯╰━━╯╰╯╰╯ ╰━━━╯ ╰━╯╰╯╰╯╰╯ ╰━━╯
\section{Carbon Stars}
\label{sec:cs}

%%%%%%%%%%%%%%%%%%%%%%%%%%%%%%%%%%%%%%%%%%%%%%%%%%%%%%%%%%%%%%%%
\subsection{Star selection}
\label{subsec:cs_star_sel}

We select the CS based on their location in the $(J-K_{s})_0$, $J_0$ CMD. As mentioned in section \ref{sec:intro}, CS are redder than the bulk of the red giants, and a colour selection is usually sufficient to separate these stars from the rest of the asymptotic giants. In Paper I we determined the colour range for the CS based on the analysis of the CS in the Magellanic Clouds. A star is catalogued as a CS if its $(J-K)_0$ colour falls within the limits:
\begin{equation}
    1.4 < (J-K_{s})_{0} < 2. 
\end{equation}
These margins reduce the contamination of oxygen-rich asymptotic giants (blue limit), and extreme CS (red limit). 

On galaxies where contamination from background galaxies is present in the CS colour range, we introduce a faint magnitude cut. As shown in section 6 of Paper II, this magnitude selection does not affect the distance estimation. 

In the Magellanic Clouds, the CS are much brighter than background galaxies, therefore it is not necessary to introduce a faint magnitude cut. For our target galaxies, we are able to remove most of the background contamination from the CMD, nevertheless some background galaxies remain. 

\begin{figure*}
\centering
\includegraphics[width=\textwidth]{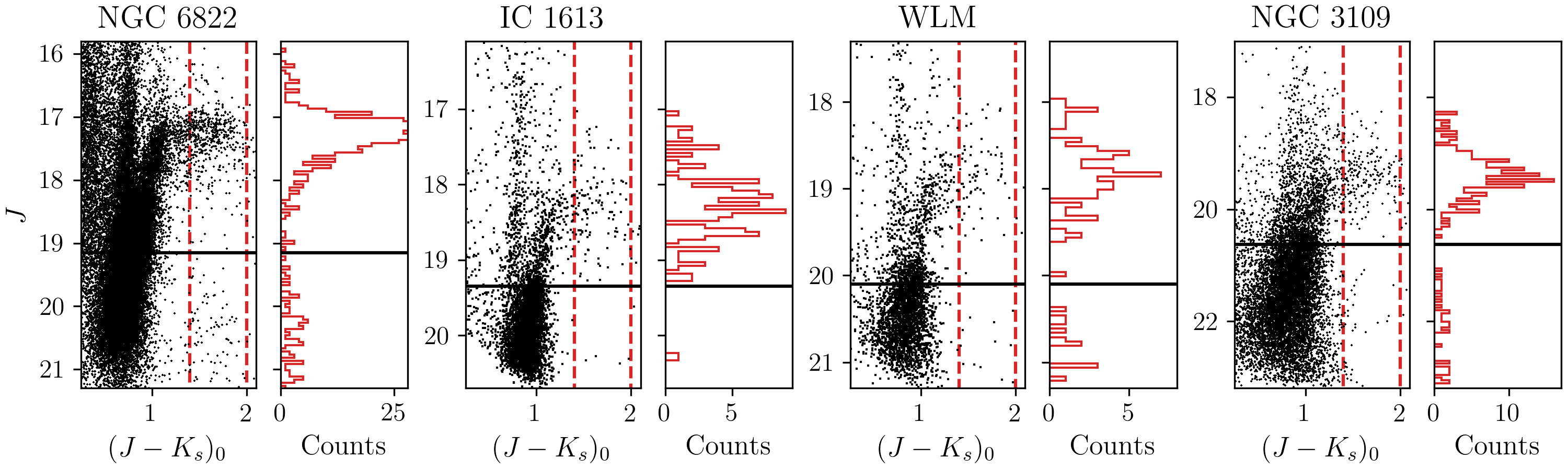}
\caption{$(J-K_{s})_0$, $J_0$ colour-magnitude diagrams (CMD) for (from left to right) NGC 6822, IC 1613, WLM, and NGC 3109. The red dashed lines represent the limits of the carbon stars selection in colour (1.4 $< (J-K)_0 <$ 2.0). The panels to the right of each CMD shows the distribution of stars within the colour range defined for the selection of the carbon stars. The black horizontal lines in each plot mark the faint magnitude limit for the selection of carbon stars in each galaxy.  }
\label{fig:CSsel}
\end{figure*}

Even though the inclusion or exclusion of these non-member sources does not disrupt the determination of the distance, they do impact the model fitting to the luminosity function of the CS. \erased{The faint $J_0$ magnitude limits for our target galaxies are: 19.15 for NGC 6822, 19.3 for IC 1613 (for both data sets), 20.2 for WLM, and 20.6 for NGC 3109.} 
Figure \ref{fig:CSsel} shows the CMD with the CS colour limits and faint magnitude cut for each galaxy in our sample. For NGC 6822 there is a significant number of faint sources that results in a small bump in the luminosity function below magnitude $\sim20$. A relatively flat distribution of stars can be seen between the two main features of the luminosity function. We place the faint magnitude cut for this galaxy at roughly the centre of this flat region at $J_0$=19.15. For IC 1613 (WIRCam), WLM and NGC 3109, the remaining contamination at the faint end of the CS region is minimal. Nevertheless, we remove the few faint sources to avoid any interference with the luminosity function fitting procedure. The faint $J_0$ magnitude limits are 19.3 for IC 1613 (for both data sets), 20.2 for WLM, and 20.6 for NGC 3109.

%%%%%%%%%%%%%%%%%%%%%%%%%%%%%%%%%%%%%%%%%%%%%%%%%%%%%%%%%%%%%%%%
\subsection{Luminosity function}
\label{subsec:cs_lum_func}

As detailed in section 3.2 of Paper II, we model the carbon star luminosity function using a modified Lorentzian model that allows for asymmetry (skewness) and different weights on the tails (kurtosis):
\begin{equation}
    f(J;m,w,s,k) = \frac{a}{1 + \left( \frac{J-m}{w} \right)^2 + s\left( \frac{J-m}{w} \right)^3 + k\left( \frac{J-m}{w} \right)^4},
\label{eq:lfmodel}
\end{equation}
where:
\begin{itemize}[leftmargin=*,label={}]
    \item $m$ = mode of the distribution,
    \item $w$ = scale parameter (specifies the width),
    \item $s$ = skewness parameter,
    \item $k$ = kurtosis parameter,
    \item $a$ = amplitude (height of the peak).
\end{itemize}

In Paper II we obtained the best-fit model using non-linear least squares fitting. In this paper we include a new parameter estimation method using an un-binned maximum likelihood estimator (MLE). The likelihood function is given by:
\begin{equation}
    \ln(L) = n \times \ln\left(\frac{1}{N}\right) + \sum_{i=1}^n \ln(f_i),
\label{eq:lfmaxlike}
\end{equation}
where $f$ is given by equation \ref{eq:lfmodel}, $n$ is the number of stars, and $N$ is given by:
\begin{equation}
    N = \int_{J_\textrm{bright}}^{J_\textrm{faint}} f(J|m,w,s,k) \quad dJ.
\end{equation}

The main difference between the application of the two fitting methods is how the data are used. For the least squares fitting the data are binned while the MLE fits the data directly without the need of binning. To estimate the parameter values we repeat the best-fitting procedure for both the binned and un-binned estimations over ten-thousand bootstrap realizations for each galaxy. The values and statistical errors for the luminosity function parameters from the bootstrapping procedure are summarized in Table \ref{tab:best-fit_lmcsmc} for the LMC and SMC and Table \ref{tab:best-fit_targets} for NGC 6822, IC 1613, WLM and NGC 3109. The results for the Lorentzian model parameters between the two methods agree within the error bars (except for the amplitude which is not included in the un-binned MLE estimation). An important improvement introduced by the MLE is the reduction in the error bars for the skew parameter. As we will show in the next section, this parameter is fundamental in determining the correct distance calibrator.

The parameters obtained for the LMC and SMC show a significant difference between the shapes of the luminosity functions of the CS in these two galaxies. The LMC is skewed towards brighter magnitudes while the SMC is fairly symmetric. The difference in symmetry is also noticeable in the difference between the value of the mode and the median $M_J$ magnitude: in the LMC these two values differ by almost a tenth of a magnitude while in the SMC the values of the median and the mode are almost identical. We see no significant difference between the values obtained from the binned and un-binned techniques.

% For our target galaxies the skew parameters shows that NGC 6822 is skewed towards brighter magnitudes as the LMC while IC 1613, WLM and NGC 3109 are more symmetric resembling the SMC. 
For NGC 6822, even though the difference in the skewness values obtained from the binned and MLE methods falls slightly outside the error bars, both results indicate that the luminosity function is significantly skewed towards brighter magnitudes, similar to that for the LMC. For IC 1613 (both data sets), WLM, and NGC 3109 the luminosity functions are more symmetric as for the SMC. As is the case for the LMC and SMC, for all four of these galaxies, the difference between the mode and median values reflects the value of the skewness for each distribution, with symmetric luminosity functions having mode$\approx$median. In the case of IC 1613, we also compare the results for the new WIRCam data set with the Sibbons catalogue used in Paper II. We find no significant difference between the values of parameters obtained from the two different catalogues. The mode for the WIRCam data is slightly fainter than that of the Sibbons data, but it remains within the error bars. %We find a significant difference in the value of the median $J_0$ magnitude with the WIRCam data having a median $\sim0.1$ magnitudes fainter compared to the Sibbons catalogue. Further discussion of this difference and its implications will be presented in the following section.

\begin{figure*}
\centering
\includegraphics[width=\textwidth]{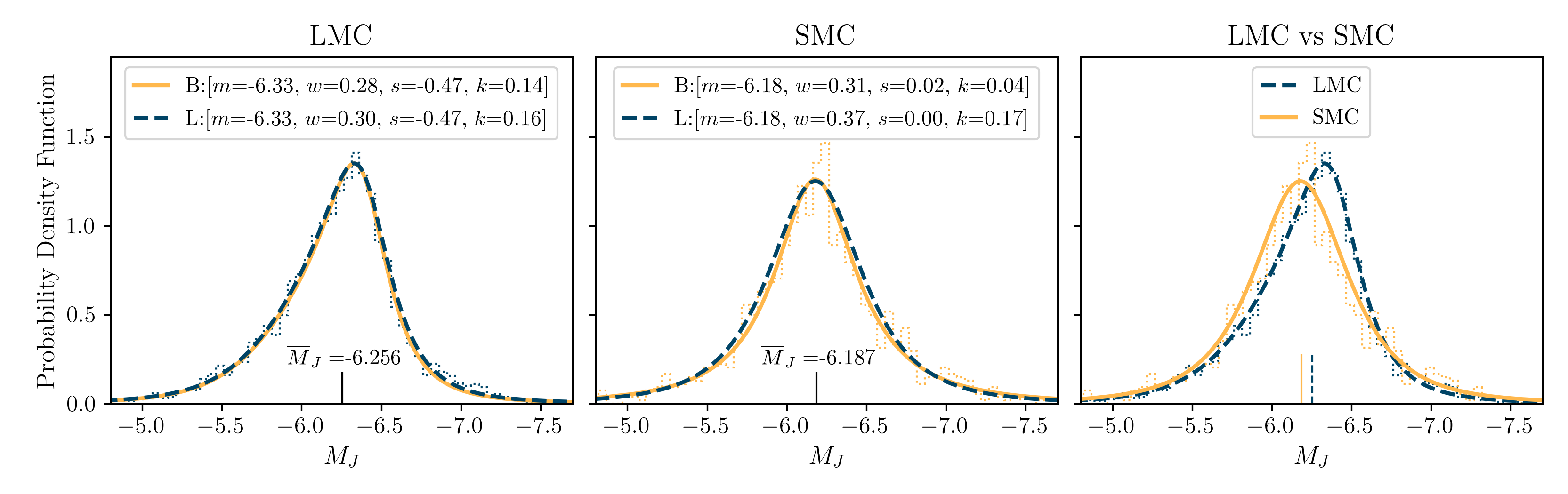}
\caption{Best-fit modified Lorentzian function for the carbon star luminosity functions for the LMC (\textit{left} panel) and SMC (\textit{centre} panel). The histograms behind each model curve depict the actual data for each galaxy. For the \textit{left} and \textit{centre} panels the straight lines represent the best-fit model using using non-linear least squares fitting to the binned data (B), while the dashed-line represents the model obtained through an un-binned maximum likelihood estimator (L). The inset specifies the values for the parameters for the Lorentzian function specified in Eq. \ref{eq:lfmodel} for both fitting methods. The vertical lines at the bottom of the plots indicate the location of the median $M_J$ magnitude with its error bars. The \textit{right} panel shows the LMC (dashed line) and SMC (straight line) L luminosity functions superimposed over each other for visual comparison. It is evident that the luminosity functions differ from one another especially in the skew parameter with the LMC skewed towards brighter magnitudes.}
\label{fig:best-fit_mc}
\end{figure*}

\begin{figure*}
\centering
\includegraphics[width=\textwidth]{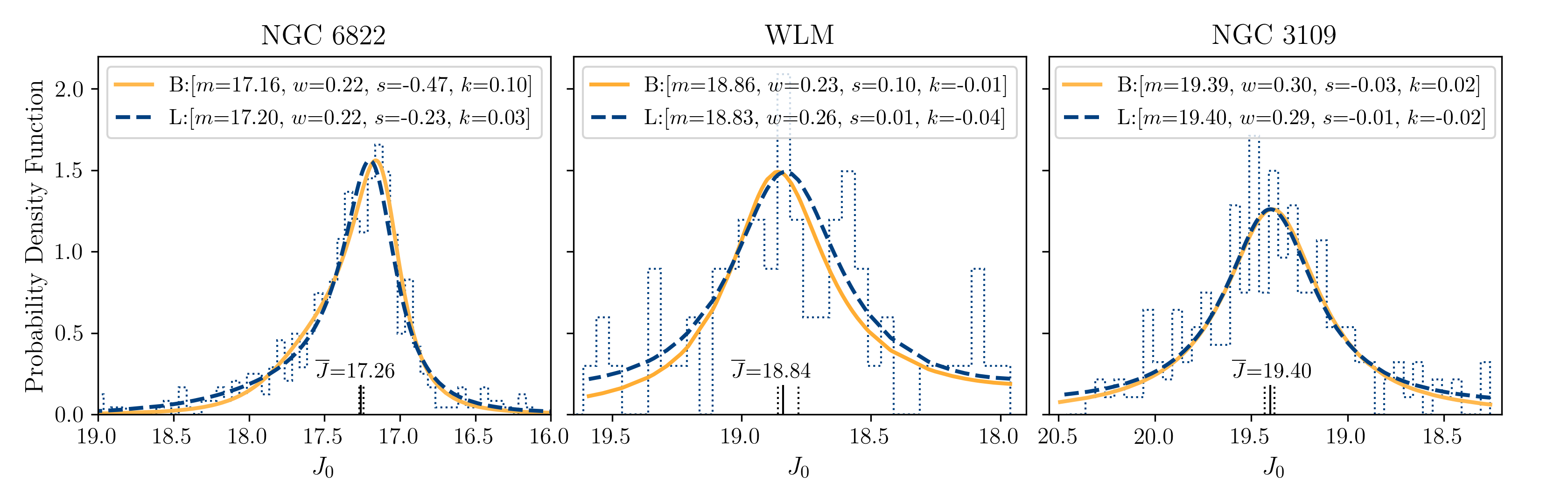}
\caption{Best-fit modified Lorentzian function for the carbon star luminosity functions for NGC 6822 (\textit{left} panel), WLM (\textit{centre} panel), and NGC 3109 (\textit{right} panel). The solid orange lines represent the best-fit model using using non-linear least squares fitting to the binned data (B), while the dark dashed-line represents the model obtained through an un-binned maximum likelihood estimator (L). The inset specifies the values for the parameters for the Lorentzian function specified in Eq. \ref{eq:lfmodel} for both fitting methods. The vertical lines at the bottom of the plots indicate the location of the median $J_0$ magnitude with its error bars.}
\label{fig:best-fit_targets}
\end{figure*}

\begin{figure*}
\centering
\includegraphics[width=\textwidth]{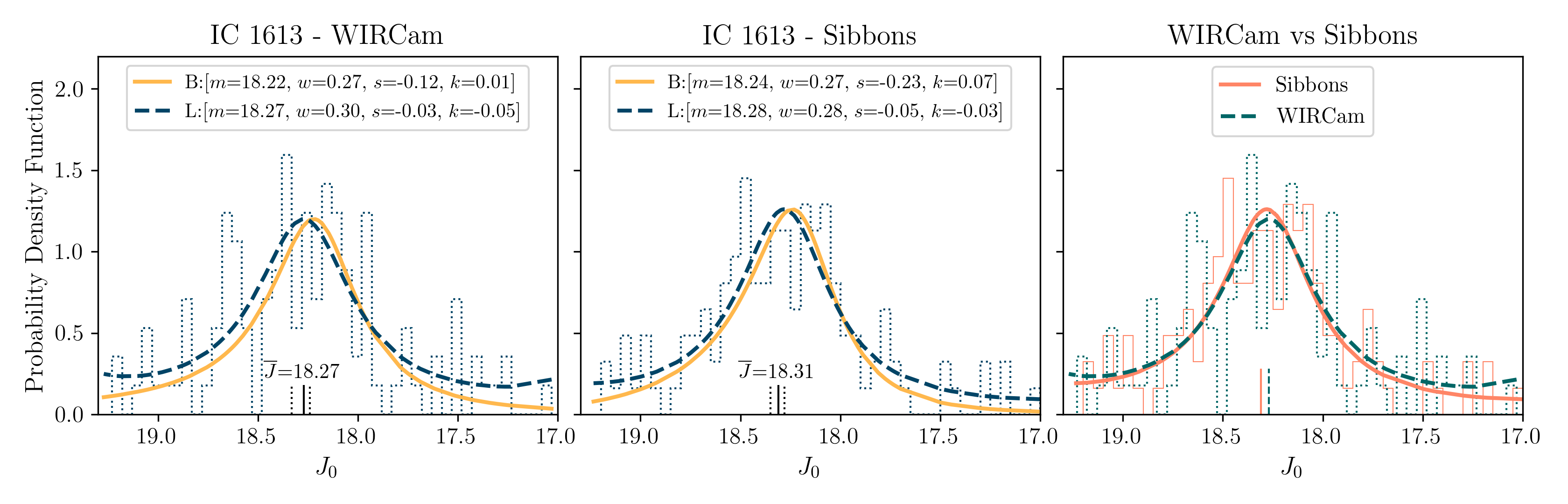}
\caption{Similar to Fig. \ref{fig:best-fit_mc} but comparing the best-fit result for the two different data sets available for IC 1613: the WIRCam data (\textit{left} panel), and the Sibbons catalogue (\textit{centre} panel). The \textit{right} panel shows the luminosity functions obtained from the maximum likelihood method superimposed over each other for visual comparison.}
\label{fig:best-fit_ic1613}
\end{figure*}

The luminosity functions along with the the best-fit models are plotted in Fig. \ref{fig:best-fit_mc} for the Magellanic Clouds, Fig. \ref{fig:best-fit_targets} for NGC 6822, WLM and NGC 3109, and Fig. \ref{fig:best-fit_ic1613} for IC 1613.

% TABLE: Bootstrap results LMC & SMC
\begin{table}
\caption{Large and Small Magellanic Clouds median absolute $J$ magnitudes ($\overline{M}_{J}$) and best-fitting parameters for the CS luminosity function model specified in Eq. \ref{eq:lfmodel}. Values and errors are obtained from 10000 bootstrap realizations.}
\centering
\begin{tabular}{l r r r r}
\hline
& \multicolumn{2}{c}{LMC} & \multicolumn{2}{c}{SMC} \\
\hline
\hline
% \vspace{0.1cm}
$\overline{M}_{J}$  & \multicolumn{2}{c}{-6.256$^{+0.005}_{-0.005}$} & \multicolumn{2}{c}{-6.187$^{+0.012}_{-0.014}$}  \\
\hline
& \multicolumn{1}{c}{Binned} & \multicolumn{1}{c}{Max.Like.} & \multicolumn{1}{c}{Binned} & \multicolumn{1}{c}{Max.Like.} \\
\hline
\vspace{0.16cm}
Mode & -6.33$^{+0.01}_{-0.01}$ & -6.33$^{+0.01}_{-0.01}$ & -6.18$^{+0.02}_{-0.02}$ & -6.18$^{+0.01}_{-0.01}$\\
\vspace{0.16cm}
Width &  0.28$^{+0.01}_{-0.01}$ & 0.30$^{+0.01}_{-0.01}$ & 0.31$^{+0.03}_{-0.03}$ & 0.37$^{+0.03}_{-0.03}$ \\
\vspace{0.16cm}
Skew & -0.47$^{+0.05}_{-0.05}$ & -0.47$^{+0.04}_{-0.03}$ & 0.02$^{+0.08}_{-0.06}$ & 0.00$^{+0.001}_{-0.002}$ \\
\vspace{0.16cm}
Kurtosis &  0.14$^{+0.03}_{-0.02}$ & 0.16$^{+0.02}_{-0.02}$ & 0.04$^{+0.05}_{-0.02}$ & 0.17$^{+0.09}_{-0.06}$ \\
Amplitude &  1.35$^{+0.03}_{-0.03}$ & & 1.26$^{+0.06}_{-0.06}$ &  \\
\hline
\end{tabular}
\label{tab:best-fit_lmcsmc}
\end{table}
% 

% TABLE: Bootstrap results target galaxies
\begin{table*}
 \caption{Similar to Table \ref{tab:best-fit_lmcsmc} for the target galaxies but with the apparent median $J$ magnitudes ($\overline{J_0}$).}
 \centering
 \begin{tabular}{l r r r r r r r r r r}
 \hline
 & \multicolumn{2}{c}{NGC 6822} & \multicolumn{2}{c}{IC 1613 - WIRCam} & \multicolumn{2}{c}{IC 1613 - Sibbons et al.}  & \multicolumn{2}{c}{WLM} & \multicolumn{2}{c}{NGC 3109} \\
 \hline
 \hline
% \vspace{0.16cm}
$\overline{J_0}$ & \multicolumn{2}{c}{17.26$^{+0.03}_{-0.02}$} & \multicolumn{2}{c}{18.27$^{+0.06}_{-0.03}$} &
\multicolumn{2}{c}{18.31$^{+0.04}_{-0.03}$} & \multicolumn{2}{c}{18.84$^{+0.02}_{-0.06}$} & \multicolumn{2}{c}{19.40$^{+0.03}_{-0.02}$} \\
\hline
& \multicolumn{1}{c}{Binned} & \multicolumn{1}{c}{Max.Like.} &
\multicolumn{1}{c}{Binned} & \multicolumn{1}{c}{Max.Like.} & \multicolumn{1}{c}{Binned} & \multicolumn{1}{c}{Max.Like.} & \multicolumn{1}{c}{Binned} & \multicolumn{1}{c}{Max.Like.} & \multicolumn{1}{c}{Binned} & \multicolumn{1}{c}{Max.Like.} \\
\hline
\vspace{0.16cm}
Mode &  17.16$^{+0.03}_{-0.03}$ & 17.20$^{+0.02}_{-0.02}$ & 18.22$^{+0.09}_{-0.09}$ & 18.27$^{+0.05}_{-0.05}$ & 18.24$^{+0.20}_{-0.13}$ & 18.28$^{+0.05}_{-0.05}$ & 18.86$^{+0.07}_{-0.06}$ & 18.83$^{+0.06}_{-0.06}$ & 19.39$^{+0.05}_{-0.06}$ & 19.40$^{+0.03}_{-0.04}$ \\
\vspace{0.16cm}
Width &   0.22$^{+0.04}_{-0.03}$ & 0.22$^{+0.02}_{-0.02}$ & 0.27$^{+0.10}_{-0.12}$ & 0.30$^{+0.08}_{-0.06}$ & 0.27$^{+0.11}_{-0.14}$ & 0.28$^{+0.06}_{-0.04}$ & 0.23$^{+0.08}_{-0.09}$ & 0.26$^{+0.09}_{-0.06}$ & 0.30$^{+0.08}_{-0.06}$ & 0.29$^{+0.07}_{-0.05}$ \\
\vspace{0.16cm}
Skew &  -0.47$^{+0.12}_{-0.23}$ & -0.24$^{+0.08}_{-0.08}$ &  -0.12$^{+0.14}_{-0.40}$ & -0.03$^{+0.04}_{-0.06}$ & -0.23$^{+0.69}_{-0.42}$ & -0.05$^{+0.05}_{-0.08}$ &  0.10$^{+0.19}_{-0.34}$ & 0.01$^{+0.06}_{-0.02}$  & -0.03$^{+0.11}_{-0.35}$ & -0.01$^{+0.03}_{-0.10}$ \\
\vspace{0.16cm}
Kurtosis &  0.10$^{+0.16}_{-0.05}$ & 0.03$^{+0.02}_{-0.02}$  &  0.01$^{+0.09}_{-0.07}$  & -0.05$^{+0.02}_{-0.02}$  & 0.07$^{+0.19}_{-0.08}$ & -0.03$^{+0.03}_{-0.01}$ & -0.01$^{+0.07}_{-0.05}$ & -0.04$^{+0.03}_{-0.04}$  & 0.02$^{+0.16}_{-0.05}$  & -0.02$^{+0.07}_{-0.02}$ \\

Amplitude & 1.56$^{+0.14}_{-0.11}$  &  & 1.20$^{+0.31}_{-0.18}$ &  & 1.26$^{+0.32}_{-0.18}$  &  & 1.49$^{+0.53}_{-0.28}$ &  & 1.26$^{+0.17}_{-0.14}$ &  \\
 \hline
 \end{tabular}
 \label{tab:best-fit_targets}
\end{table*}

%%%%%%%%%%%%%%%%%%%%%%%%%%%%%%%%%%%%%%%%%%%%%%%%%%%%%%%%%%%%%%%%
\subsection{Distance Determination Method}
\label{subsec:cs_dist_det}

Our distance determination method is based on the standard candle aspect of the median $J_0$ magnitude of the carbon stars. This attribute of the carbon stars and how it might be affected by the intrinsic properties of the host galaxy are discuss at length in Paper II (section 3 and 6). We choose the median over the mean $J_0$ magnitude, as the median is less sensitive to outliers and to the effects of magnitude selection (such as the faint magnitude limits set for our target galaxies). 

To establish the distance to each galaxy we first need to determine the most suitable calibrator between the LMC and SMC. To classify a galaxy into "LMC-like" or "SMC-like" we use the shape of the CS luminosity function as described by the best-fit parameters defined in section \ref{subsec:cs_lum_func}, specifically the skew value. The skew parameter tells us how symmetric or asymmetric is the luminosity function. For example, a negative skewness indicates the luminosity function is inclined towards brighter magnitudes. Given a galaxy $X$ with skewness $s_{X}$ the distance modulus is defined by the difference between then median $J_0$ magnitude (${\overline{J}_0}$) of $X$ and the median of the absolute $J$ magnitude ($\mathrm{\overline{M}}_{J}$) of the calibrator:
\begin{equation}
    \mu_{0,X} =
\begin{cases}
    \overline{J}_{0,X} - \mathrm{\overline{M}}_{J_{LMC}},& \text{if } s_{X} < -0.20 \vspace{0.2cm} \\ 
    \overline{J}_{0,X} - \mathrm{\overline{M}}_{J_{SMC}},& \text{otherwise},
\end{cases}
\label{eq:dm}
\end{equation}
where $-0.20$ is approximately midway between the skewness of the two calibrators (the LMC and SMC). The absolute magnitudes for the Magellanic Clouds are set by adopting $\mu_{\mathrm{LMC}}$ = 18.477 $\pm$ 0.004 (statistical) $\pm$ 0.026 (systematic) \citep{2019Natur.567..200P} for the LMC, and $\mu_{\mathrm{SMC}}$ = 18.977 $\pm$ 0.016 (statistical) $\pm$ 0.028 (systematic) \citep{2020ApJ...904...13G} for the SMC.

%%%%%%%%%%%%%%%%%%%%%%%%%%%%%%%%%%%%%%%%%%%%%%%%%%%%%%%%%%%%%%%%
\subsection{Results}
\label{subsec:cs_results}
Each target galaxy's CS luminosity function was fitted using two different methods: a direct fit to the binned data (refer as the "binned" method) and an un-binned maximum likelihood estimator (MLE). As discussed in section \ref{subsec:cs_lum_func} the values obtained for the model parameters using both methods agree within the error bars. In this section we use the luminosity function skew parameter value to determine the most suitable calibrator for each galaxy and estimate its distance modulus. 

\subsubsection{NGC 6822}
As in Paper II, the best-fit skew parameter indicates NGC 6822 is significantly skewed towards brighter magnitudes while its width seems narrower than the rest of the galaxies in our sample. According to the value obtained for the skewness: $s_\textrm{binned}=-0.47$ and $s_\textrm{MLE}=-0.24$ both $<-0.20$, NGC 6822 is classified as an "LMC-like" galaxy. Therefore we use the LMC as the calibrator to obtain the distance to NGC 6822. Using equation \ref{eq:dm} we obtain a true distance modulus of \inserted{$\mu_0 = 23.52 \pm 0.03$} (statistical). 

Two other independent estimates of the distance to NGC 6822 using CS have recently been published. \citet{2021ApJ...916...19Z} obtained a distance modulus of $\mu_0 = 23.243 \pm 0.037$ (stat), while \citet{2020ApJ...899...67F} reports $\mu_0 = 23.44 \pm 0.02$ (stat) (not formally in the paper but as part of the appendix). All three results are in disagreement with each other. NGC 6822 is a challenging target due to the considerable amount of reddening in the line-of-sight, in the case of the three estimates mentioned here, modifying the different $E(B-V)$ values used in each study does not bring the results into agreement. The distance modulus from \citet{2021ApJ...916...19Z} is considerably smaller than the value used in this paper and in \citet{2020ApJ...899...67F}. 

\erased{The near-infrared data employed by \citet{2021ApJ...916...19Z} are the same data used by \citet{2006ApJ...647.1056G} to determined the distance to NGC 6822 using Cepheid stars. \citet{2009ApJ...693..936M} found a systematic shift in the zero-point in the $J$ and $K$ photometry from \citet{2006ApJ...647.1056G} and stated that inclusion of the $J$ and $K$ data in their analysis, would have resulted in a distance modulus 0.17 mag brighter. \cite{2020ApJ...899...67F} compile $I$-band $\mu_0(TRGB)$ results from the literature. To take an average the independent estimates are standardize with $A_I=0.355$ and $M_I^{TRGB}=-4.05$, giving an average of $\mu_0(TRGB)=23.52$, in good agreement with our CS result.}

\textnew{A more detailed discussion of the effects of different extinction values and a comparison between our results and other independently measured distances to NGC 6822 is presented in Sec. \ref{sec:discussion}. }

\subsubsection{IC 1613}

The results obtained from the \cite{2015A&A...573A..84S} catalogue differ from those obtained in Paper II reflecting the corrections explained in Sec. \ref{sec:ic1613sibdata}. Our analysis of IC 1613 in Paper II had over-corrected for reddening and we had erroneously neglected the transformation between UKIRT/WFCAM and 2MASS colours. The results for the luminosity function fitting parameters (mode, width, skewness and kurtosis) are in very good agreement between the data sets for both the binned non-linear square fitting and the un-binned maximum likelihood estimator methods. The one parameter that stands out is the skewness obtained for the Sibbons data using the binned method, which seems to be lower than expected. Nevertheless, we have to consider that the error bars for this specific case are significantly larger than in all other skew parameter results. 

To classify IC 1613 we consider the skewness values obtained for both data sets, except for the binned estimation carried out on the Sibbons catalogue given the large uncertainty in the result. Considering the other three skewness values obtained for IC 1613, this galaxy is catalogued as an "SMC-like" galaxy. Using the SMC as calibrator we obtain \inserted{$\mu_0 = 24.50 \pm 0.04$} (statistical) and \inserted{$\mu_0 = 24.46 \pm 0.05$} (statistical) for the Sibbons and WIRCam data respectively. The distance moduli obtained from the two different data sets are in good agreement within the error bars.

Only one other estimate of the distance modulus to IC 1613 using CS has been perform to date. \cite{2020ApJ...899...67F} report $\mu_0 = 24.36 \pm 0.03$ (stat). This result is not consistent with the distance modulus found in this work, and the statistical errors are not enough to bring them into agreement. 

\erased{Recently, \cite{2021arXiv211106899N} used Gaia EDR3 data to estimate distances to several dwarf galaxies based on RR Lyrae variable stars. In their study, \citeauthor{2021arXiv211106899N} report $\mu_0 = 24.42 \pm 0.03$ for IC 1613. This value is in agreement with our results within the statistical error bars. \citeauthor{2021arXiv211106899N} also do an extensive comparison with other distance estimates for IC 1613 published over the last decade. They find that pre-2014 estimates are closer to $\mu \sim 24.45$, compared to $\mu \sim 24.3$ for more recent estimates. }

\subsubsection{WLM \& NGC 3109}
WLM and NGC 3109 show highly symmetric CS luminosity functions with skew values of $s_{binned}=-0.10$ and $s_{MLE}=0.01$ for WLM and $s_\textrm{binned}=-0.03$ and $s_\textrm{MLE}=-0.01$ for NGC 3109. All these values are greater than the limit specified in equation \ref{eq:dm}, therefore, both galaxies are classified as "SMC-like" galaxies. Using the most suitable calibrator for these two galaxies we obtain distance moduli of \inserted{$\mu_0 = 25.03 \pm 0.04$} (statistical) for WLM and \inserted{$\mu_0 = 25.59 \pm 0.03$} (statistical) for NGC 3109.

Two other studies using CS report a distance modulus for NGC 3109 and three for WLM. \cite{2021ApJ...916...19Z} obtained a distance modulus of $\mu_0 = 24.954 \pm 0.08$ (stat) for WLM and $\mu_0 = 25.517 \pm 0.046$ (stat) for NGC 3109, while \cite{2020ApJ...899...67F} reports $\mu_0 = 24.97 \pm 0.05$ (stat) and $\mu_0 = 25.56 \pm 0.05$ (stat) for WLM and NGC 3109 respectively. In the case of \cite{2021ApJ...907..112L} they find a distance modulus to WLM of $\mu_0 = 24.97 \pm 0.02$ (stat). The results reported in the current study and the papers cited agree within the error bars. 

\begin{figure*}
\centering
\includegraphics[width=\textwidth]{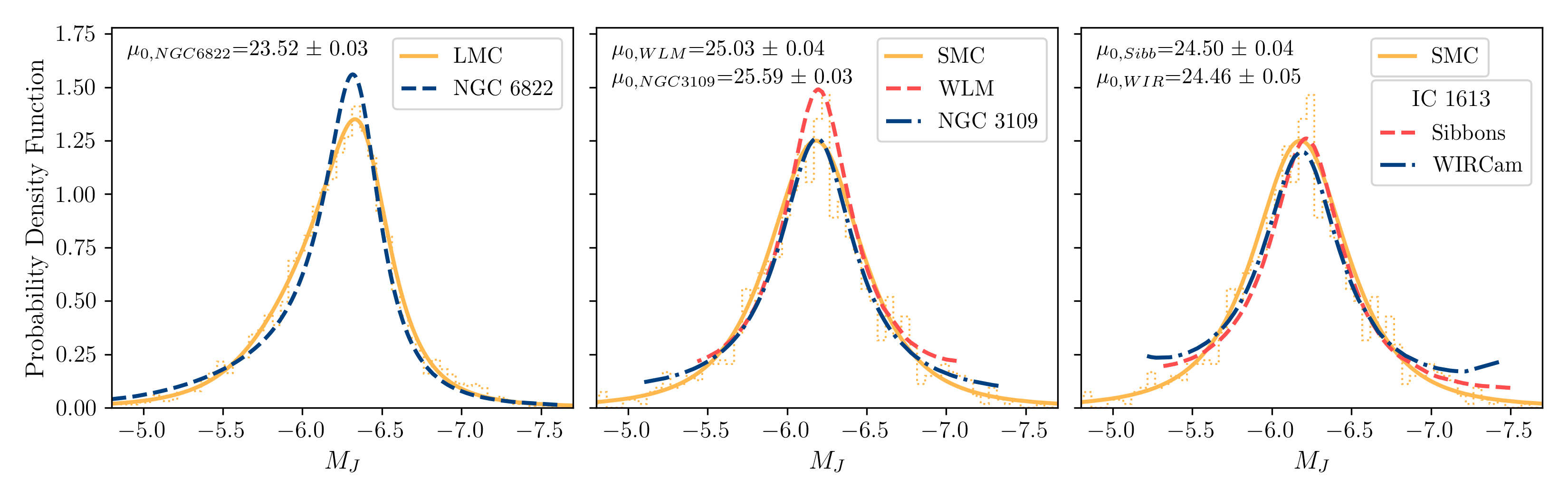}
\caption{Carbon star luminosity functions of the target galaxies plotted over the luminosity function of the most suitable calibrator. In the \textit{left} panel we display the LMC (solid orange line) and NGC 6822 (dashed blue line). In the \textit{centre} panel is the SMC (solid orange line) together with WLM (dashed light red line), and NGC 3109 (dashed-dotted blue line). In the \textit{right} panel we show the results for IC 1613 from the two different data sets available plotted over the SMC. The apparent $J_0$ magnitude distributions for the target galaxies from Figs. \ref{fig:best-fit_targets} and \ref{fig:best-fit_ic1613} have been transformed to absolute magnitudes using the distance modulus ($\mu_0$) specified in each plot. The errors shown for the distance moduli are the statistical errors.}
\label{fig:lumfunc_dm}
\end{figure*}

Fig. \ref{fig:lumfunc_dm} summarizes the results presented above. The CS luminosity function for each galaxy is shifted by its distance modulus and plotted over the CS luminosity function of its most suitable calibrator. The plots in Fig. \ref{fig:lumfunc_dm} allow us to visually compare the shape of the distributions and to see how they resemble the luminosity functions of the calibrators. 

%%%%%%%%%%%%%%%%%%%%%%%%%%%%%%%%%%%%%%%%%%%%%%%%%%%%%%%%%%%%%%%%
%%%%%%%%%%%%%%%%%%%%%%%%%%%%%%%%%%%%%%%%%%%%%%%%%%%%%%%%%%%%%%%%
%%%%%%%%%%%%%%%%%%%%%%%%%%%%%%%%%%%%%%%%%%%%%%%%%%%%%%%%%%%%%%%%
% ╭━━━╮      ╭╮ ╭━━━╮          ╭╮
% ┃╭━╮┃      ┃┃ ┃╭━╮┃         ╭╯╰╮
% ┃╰━╯┃╭━━╮╭━╯┃ ┃┃ ╰╯╭╮╭━━╮╭━╮╰╮╭╯╭━━╮
% ┃╭╮╭╯┃ ━┫┃╭╮┃ ┃┃╭━╮╭╮┃╭╮┃┃╭╰╮┃┃ ┃ ━┫
% ┃┃┃╰╮┃ ━┫┃╰╯┃ ┃╰┻ ┃┃┃┃╭╮┃┃┃┃┃┃╰╮┣━ ┃
% ╰╯╰━╯╰━━╯╰━━╯ ╰━━━╯╰╯╰╯╰╯╰╯╰╯╰━╯╰━━╯
\section{The tip of the red giant branch}
\label{sec:rg}

The TRGB marks the upper end of the RGB which, in the optical and near-infrared CMDs, can be interpreted as a boundary between the RGB and the oxygen-rich AGB. The discontinuity observed in the CMDs is a consequence of the sudden drop in luminosity caused by the onset of helium burning in the core of low mass stars (stars with masses $\approx1.0-2.3 M_\odot$ \citep{2012sse..book.....K, 2017A&A...606A..33S}. The helium ignition occurs when the degenerate helium core reaches high enough temperature ($\sim10^8$ K) to fuse helium into carbon and oxygen via the triple-$\alpha$ process. This explosive event changes the structure of the star from a red giant to a horizontal branch star in a time much shorter than the dynamical time scale, making the star suddenly bluer and less luminous. Because the helium ignition occurs at a specific temperature, the magnitude of the TRGB can be predicted and used as a standard candle (see \cite{2018SSRv..214..113B} and references therein). 

Studies of the TRGB behaviour in the near-infrared were scarce but the development of JWST spiked interest in these wavelengths. \cite{2004MNRAS.354..815V} derived a metallicity-based calibration based on globular clusters. Using these metallicity relationships \cite{2016AJ....151..167G} derived a colour-based calibration for the $J$ and $K_s$ filters. \cite{2018ApJ...858...11M} and \cite{2018ApJ...858...12H} calibrated the TRGB using IC 1613 and the LMC respectively. Their results were in good agreement with those found theoretically by \cite{2017A&A...606A..33S}. A recent analysis on the calibration of the TRGB in the near-infrared can also be found in \cite{2020ApJ...891...57F}.

%%%%%%%%%%%%%%%%%%%%%%%%%%%%%%%%%%%%%%%%%%%%%%%%%%%%%%%%%%%%%%%%
\subsection{Red giant selection}
\label{subsec:rg_star_sel}

For the TRGB detection method, instead of using the $J_0$ or $(K_s)_0$ magnitudes directly, we use the composite magnitude $T$ introduced by \cite{2009ApJ...690..389M} and calibrated for the near-infrared $JHK_s$ bands by \cite{2018ApJ...858...11M} and \cite{2018ApJ...858...12H}. The $T$ magnitude is meant to remove the effects of metallicity on the magnitude of the TRGB. In the CMD, metallicity effects can be seen in the changing slope of the TRGB with colour, in the red or blue direction depending on the observing bands. In the near-infrared filters, we see that stars at the bluer end of the TRGB are slightly fainter than at the redder end. The slope of the TRGB can be estimated and used to rectify it. The $T$ magnitude is constructed in an attempt to rectify the TRGB by correcting each star's magnitude based on its colour and the slope of the TRGB, bringing the previously inclined TRGB to a leveled discontinuity between the RGB and the AGB.

\begin{figure}
\centering
\includegraphics[trim={1.45cm 0cm 13cm 1cm},clip,width=\columnwidth]{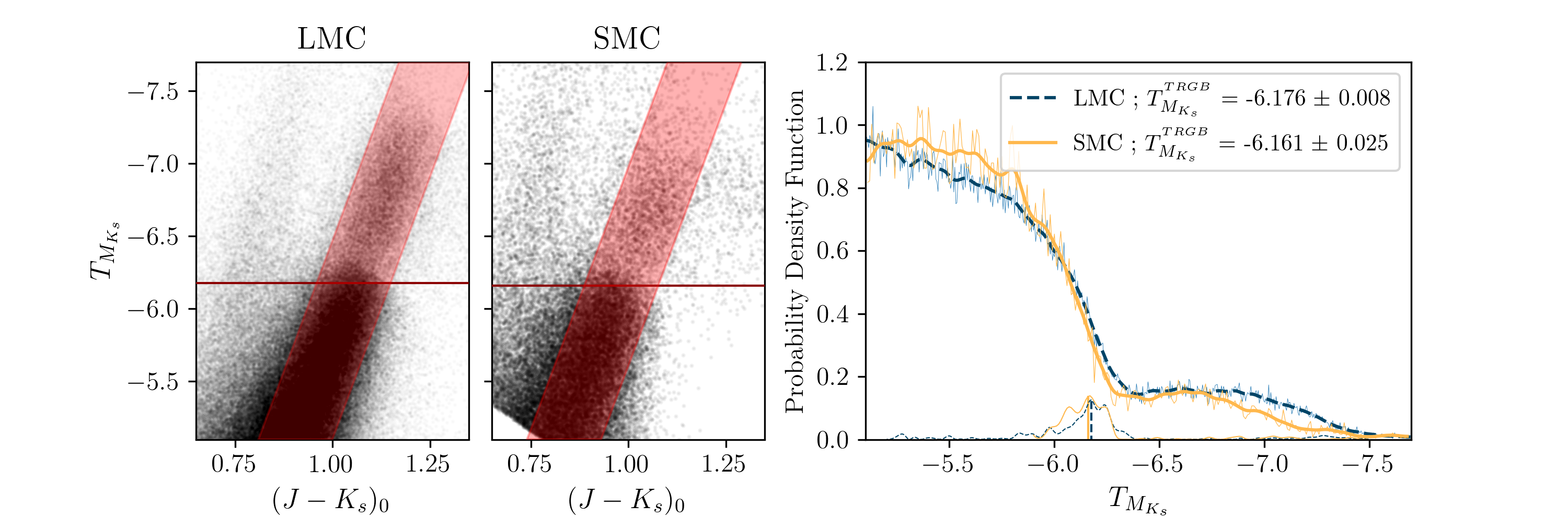}
\caption{$(J-K_{s})_0$, $T_{M_{K_s}}$ colour-magnitude diagrams for the LMC (\textit{left} panel) and the SMC (\textit{right} panel). The red-shaded area demarcates the region for the red giants selection. The horizontal red line indicates the location of the tip of the red giant branch.}
\label{fig:trgb_mc_cmds}
\end{figure}

\begin{figure*}
\centering
\includegraphics[width=\textwidth]{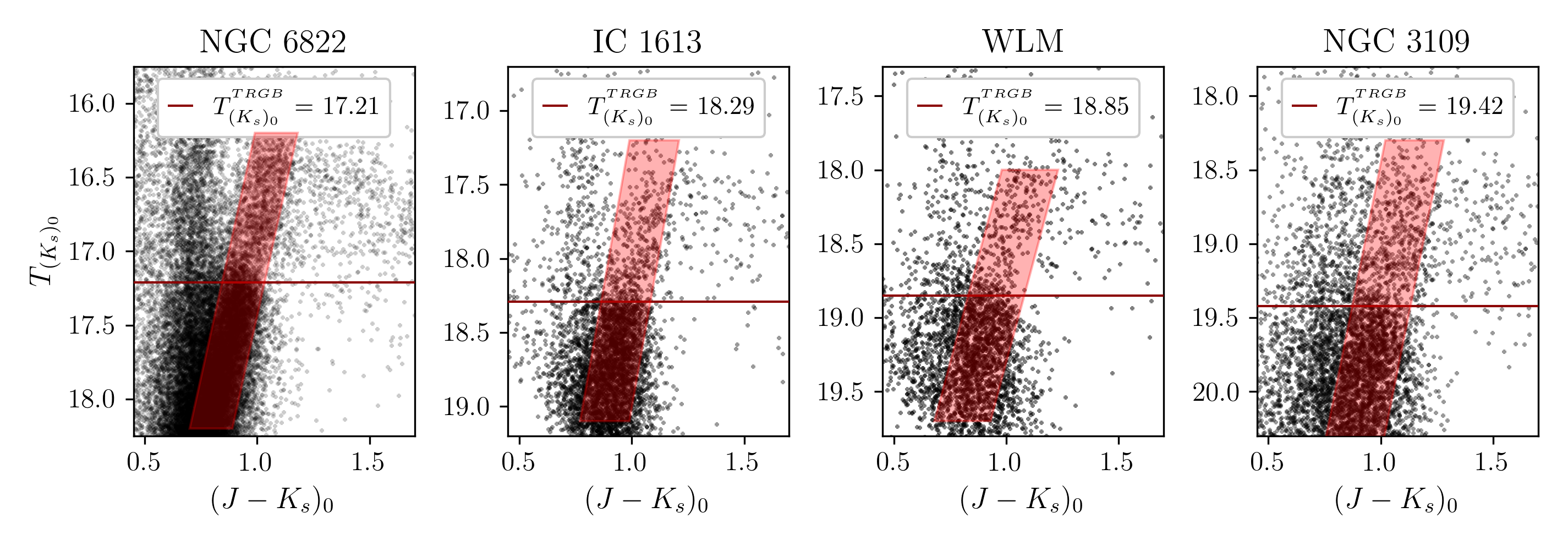}
\caption{$(J-K_{s})_0$, $T_{(K_s)_0}$ colour-magnitude diagrams for our target galaxies. The red-shaded area demarcates the region for the red giants selection. The horizontal red line indicates the location of the tip of the red giant branch whose magnitude is detailed in the insets of each panel.}
\label{fig:trgb_target_cmdK}
\end{figure*}

Based on the definition of $T$ and the slopes found in \cite{2018ApJ...858...11M} and \cite{2018ApJ...858...12H}, we assume the composite $(K_s)_0$ magnitude to be:
\begin{equation}
T_{(K_s)_0} = (K_s)_0 + 1.85[(J-K_s)_0 - 1.00].
\label{eq:tmag}
\end{equation}
In the case of the LMC and SMC, we use the absolute magnitudes $M_{K_s}$ for $(K_s)_0$. The resulting $(J-K_{s})_0$, $T$ CMDs are presented in Fig. \ref{fig:trgb_mc_cmds} for the LMC and SMC, and Fig. \ref{fig:trgb_target_cmdK} for our target galaxies.

%%%%%%%%%%%%%%%%%%%%%%%%%%%%%%%%%%%%%%%%%%%%%%%%%%%%%%%%%%%%%%%%
\subsection{TRGB detection}
\label{subsec:rg_lum_func}

\begin{figure}
\centering
\includegraphics[trim={12.6cm 0cm 2cm 0.5cm}, clip,width=\columnwidth]{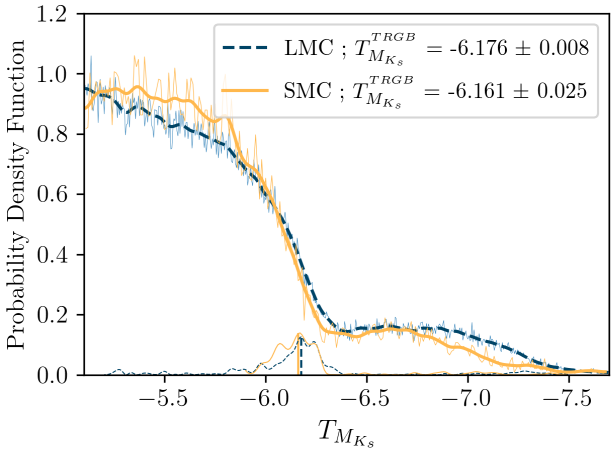}
\caption{$T_{M_{K_s}}$ red giant luminosity function. Blue-dashed lines represent the LMC and orange lines the SMC. The most prominent (thicker) lines show the smoothed luminosity functions plotted on top of the binned data (thin lines in the background). At the bottom of the plot we see the Sobel filter or edge-response functions where their maxima indicate the location of the TRGB and are marked by a vertical line of the same style and colour. The values of the TRGB for the LMC and SMC are indicated in the inset.}
\label{fig:trgb_mc_lf}
\end{figure}

\begin{figure*}
\centering
\includegraphics[width=\textwidth]{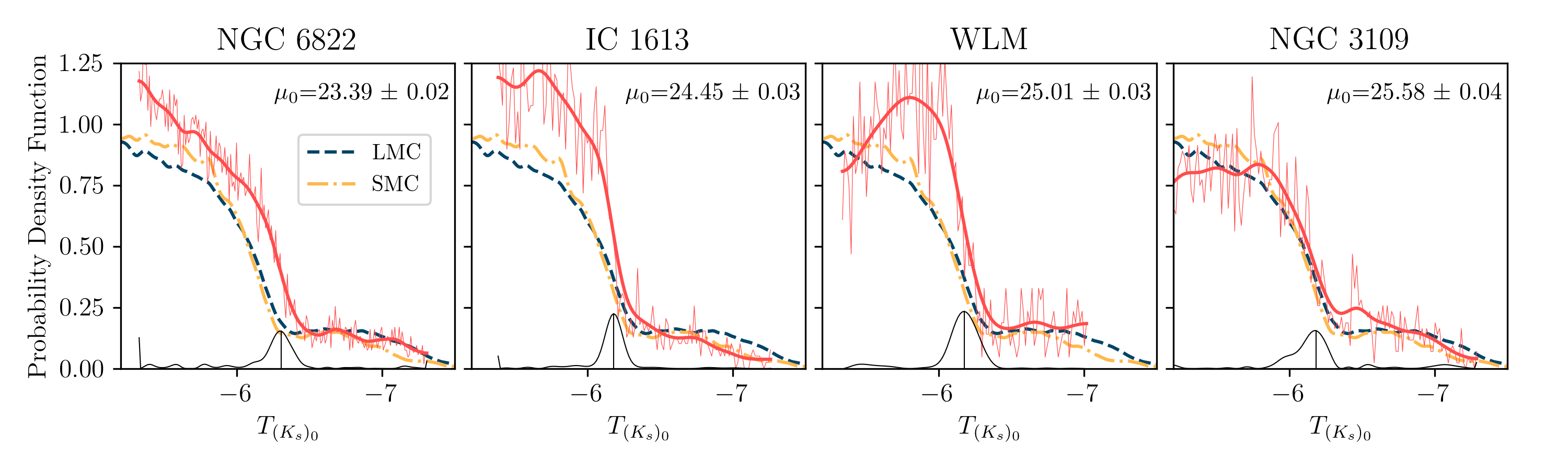}
\caption{Red giants $T_{(K_s)_0}$ luminosity functions for our target galaxies. The stars selected to draw the luminosity function are shown in Fig. \ref{fig:trgb_target_cmdK}. In each panel the red thin continuous line shows the binned luminosity function while the thicker red line on top is the smoothed luminosity function. We also plot the smoothed luminosity function of the LMC (blue dashed line) and the SMC (orange dashed-dotted line) for comparison. The black line running at the bottom of the plots shows the response of the edge-detection filter. The maximum in the response function indicates the location (i.e. magnitude) of the tip of the red giant branch. The estimated distance modulus for each galaxy is specified in each plot along with the statistical error.}
\label{fig:trgb_target_K}
\end{figure*}

We construct the luminosity function of the Magellanic Clouds and the target galaxies by taking the normalized frequency of stars in magnitude bins of size 0.005 mag for the LMC, 0.01 mag for the SMC and NGC 6822, and 0.015 mag for the IC 1613, WLM and NGC 3109. The luminosity function is then smoothed using a 1-D Gaussian filter. Once the probability density function (i.e. binned luminosity function) has been smoothed we apply a Sobel filter to detect the TRGB. This method of detection was first introduced by \cite{1993ApJ...417..553L} and 
the Sobel filter places a mask ($[-1,0,+1]$) along the smoothed luminosity function which gives the value of the first derivative at each point along the curve. Following the procedure in \cite{2021ApJ...915...34H}, we multiply the Sobel filter output by a Poisson weighting scheme: $\left(N_{i+1}-N_{i-1}\right)/\sqrt{N_{i-1}+N_{i+1}}$. This process works as an edge-response function as the maximum value of the Sobel filter marks where the slope of the luminosity function is the steepest. This maximum indicates the location of the TRGB. The binned and smoothed luminosity functions, and the Sobel filter response are shown in Fig. \ref{fig:trgb_mc_lf} for the LMC and SMC, and Fig. \ref{fig:trgb_target_K} for the target galaxies.

For each galaxy, we estimate the TRGB magnitude over ten-thousand bootstrap realizations. The values and statistical errors for the TRGB from the bootstrapping procedure are summarized in Table \ref{tab:trgb_mag_mc} for the Magellanic Clouds and Table \ref{tab:trgb_mag_target} for the target galaxies. The location of the TRGB on the CMDs are shown in Fig. \ref{fig:trgb_mc_cmds} for the LMC and SMC, and Fig. \ref{fig:trgb_target_cmdK} for the target galaxies.

\begin{table}
\caption{Tip of the red giant branch (TRGB) absolute $T$ magnitudes for the Magellanic Clouds. The results and statistical errors come from ten-thousand bootstrap realizations. The TRGB magnitudes are also shown on the colour-magnitude diagrams in Fig. \ref{fig:trgb_mc_cmds}.}
\centering
\begin{tabular}{c r r}
\hline
& \multicolumn{1}{c}{LMC} & \multicolumn{1}{c}{SMC} \\
\hline
\hline
$T_{M_{K_s}}^{TRGB}$ & -6.176$^{+0.010}_{-0.005}$  & -6.161$^{+0.030}_{-0.020}$ \\
\hline
\end{tabular}
\label{tab:trgb_mag_mc}
\end{table}

\begin{table}
\caption{Same as Table \ref{tab:trgb_mag_mc} but for the target galaxies. The TRGB magnitude are also shown on the colour-magnitude diagrams in Fig. \ref{fig:trgb_target_cmdK}.}
\centering
\begin{tabular}{l r r r r}
\hline
& \multicolumn{1}{c}{NGC 6822} & \multicolumn{1}{c}{IC 1613} & \multicolumn{1}{c}{WLM} & \multicolumn{1}{c}{NGC 3109} \\
\hline
\hline
$T_{K_{s0}}$ & 17.21$^{+0.02}_{-0.01}$ & 18.29$^{+0.01}_{-0.01}$ & 18.85$^{+0.02}_{-0.02}$ & 19.42$^{+0.03}_{-0.02}$ \\
\hline
\end{tabular}
\label{tab:trgb_mag_target}
\end{table}
% 

%%%%%%%%%%%%%%%%%%%%%%%%%%%%%%%%%%%%%%%%%%%%%%%%%%%%%%%%%%%%%%%%
\subsection{Results}
\label{subsec:rg_results}

The objective of using the $T$ magnitude is to remove possible effects of metallicity on the magnitude of the TRGB. By using this colour dependent magnitude we expect to find the absolute TRGB magnitude in each filter to be the same for the LMC and SMC. \removed{Nevertheless we find that the values obtained for the TRGB in the Magellanic Clouds show a difference of $\sim$0.11 magnitudes. Along with the difference in the absolute magnitudes of the TRGB in the LMC and SMC there are differences between the RGs luminosity functions.} \inserted{Through bootstrapping we obtained TRGB-$K_s$ magnitudes of $-6.176\pm0.008$ and $-6.161\pm0.025$ for the LMC and SMC respectively. The TRGB magnitude results for the Magellanic Clouds are in agreement within the error bars. 
The edge-detection function for the LMC shows a second peak 0.06 magnitudes brighter than the maximum response. This second peak is very near in size to the first peak and during bootstrapping it is detected $\sim30\%$ of the time. } 

When we superimpose the luminosity functions of the Magellanic Clouds we observe differences in the shapes of the $T$ magnitude luminosity functions. The most prominent difference is how steep the drop is in the number of stars between the RGB and AGB which seems more abrupt in the SMC than the LMC. Similarly, we observe different luminosity function shapes among the target galaxies. These differences in luminosity function shape and TRGB magnitude could indicate that the construction of a composite magnitude may not be enough to compensate for metallicity effects. It may also be that star formation history additionally plays an important role in defining the absolute magnitude of the TRGB. 

As with the CS, we use the TRGB for both the LMC and SMC as calibrators. To determine which calibrator is most suitable for each target galaxy we could try to determine whether these galaxies are more "LMC-like" or "SMC-like" using again their luminosity functions. Nevertheless, with only a smoothed luminosity function we cannot construct a procedure to quantify the differences between an "LMC-like" and an "SMC-like" like galaxy and the classification would be based entirely on a visual analysis of the luminosity functions.

For the purposes of this paper the determination of the distance moduli to our sample of galaxies will rely on the classification obtained from the CS analysis. From Section \ref{subsec:cs_results}, NGC 6822 was classified as an "LMC-like" galaxy, while IC 1613, WLM and NGC 3109 were classified as "SMC-like". With this classification we obtain \inserted{$\mu_0 = 23.39 \pm 0.02$ (stat) for NGC 6822, $\mu_0 = 24.45 \pm 0.03$ (stat) for IC 1613 (from the WIRCam data), $\mu_0 = 25.01 \pm 0.03$ (stat) for WLM and $\mu_0 = 25.58 \pm 0.04$ (stat) for NGC 3109}. The results obtained from the CS median $J$ magnitude and the TRGB are summarized in Table \ref{tab:dm_results}. \inserted{Except for the distance moduli obtained from the TRGB for NGC 6822,} the results obtained from both methods agree well within the statistical error bars.

%%%%%%%%%%%%%%%%%%%%%%%%%%%%%%%%%%%%%%%%%%%%%%%%%%%%%%%%%%%%%%%%

\begin{table}
\caption{Distance moduli values obtained from the median $J$ magnitude of the carbon stars compared with the distance moduli estimated using the TRGB. The errors presented are the statistical errors.}
\centering
\begin{tabular}{l r r}
\hline
& \multicolumn{1}{c}{Carbon Stars} & \multicolumn{1}{c}{TRGB} \\
\hline
\hline
\vspace{0.16cm}
NGC 6822 & 23.52 $\pm$ 0.03  & 23.39 $\pm$ 0.02    \\
\vspace{0.16cm}
IC 1613 (WIRCam)  & 24.46 $\pm$ 0.05  & 24.45 $\pm$ 0.03    \\
\vspace{0.16cm}
IC 1613 (Sibbons) & 24.50 $\pm$ 0.04  &     \\
\vspace{0.16cm}
WLM      & 25.03 $\pm$ 0.04  & 25.01 $\pm$ 0.03    \\
NGC 3109 & 25.59 $\pm$ 0.03  & 25.58 $\pm$ 0.04    \\
\hline
\end{tabular}
\label{tab:dm_results}
\end{table}
%%%%%%%%%%%%%%%%%%%%%%%%%%%%%%%%%%%%%%%%%%%%%%%%%%%%%%%%%%%%%%%%
%%%%%%%%%%%%%%%%%%%%%%%%%%%%%%%%%%%%%%%%%%%%%%%%%%%%%%%%%%%%%%%%

\section{Comparison with independent distance determinations}
\label{sec:discussion}

In this Section, the results obtained from the CS and TRGB methods are compared against previously published, independently measured distances using different distance indicators. It
is important to acknowledge the fact that a direct comparison between the values for $\mu_0$ for a given galaxy is difficult as different studies use a variety of extinction values and distance calibrations.

\subsection{NGC 6822}

Table \ref{tab:dist_n6822} displays some of the recent distance estimates to NGC 6822 using various techniques and standard candles. The method, filter and extinction values for each study are also given to allow for better comparison. The extinction value used in this work, $E(B-V)=0.35$ (equivalent to $A_J\sim0.29$ and $A_V\sim1.05$) \citep{2006ApJ...647.1056G, 2012AA...548A.129F, 2014ApJ...794..107R}, considers the total line-of-sight reddening (from the Milky Way and NGC 6822 internal reddening). On the other hand, some studies find smaller total extinction values or consider only the foreground reddening which can lead to $E(B-V)$ values as low as $\sim0.2$ (e.g. \citep{2011ApJ...737..103S, 2012AA...548A.129F, 2021ApJ...923..232R}). The large range of extinction values alone carries a systematic uncertainty of $\sim0.15$ mag in distance modulus using $J$-band and $\sim0.45$ mag in $V$-band. 

 \begin{table}
    \centering
    \caption[Independent distance estimations to NGC 6822]{Independent distance determinations to NGC 6822.} 
 \begin{tabular}{@{\hskip 0pt} c@{\hskip 8pt} c@{\hskip 7pt} c@{\hskip 7pt} c@{\hskip 7pt} c@{\hskip 0pt}}
\midrule 
Method & Band & E$(B-V)$ & $\mu_0^*$ & Reference \\
\midrule 
\midrule
   Cepheids & near-IR & (3)  & $23.43\pm0.06$    & \citet{2016AJ....151...88B}   \\
   Cepheids & (1)     & 0.35 & $23.38\pm0.02$    & \citet{2014ApJ...794..107R}   \\ 
   Cepheids & (2)     & (4)  & $23.40\pm0.05$    & \citet{2012MNRAS.421.2998F}   \\
   Cepheids & mid-IR  & 0.26 & $23.49\pm0.03$    & \citet{2009ApJ...693..936M}   \\
   Cepheids & $JK$ & 0.356 & $23.29\pm0.02^{\dag}$ & \citet{2006ApJ...647.1056G} \\
\midrule
  TRGB & $K_s$ & 0.35 & 23.39 $\pm$ 0.02 & This work \\
  TRGB & $I$ & (5)  & $23.52\pm0.03^{\ddag}$ & F\&M \citeyear{2020ApJ...899...67F}\\
  TRGB & $I$ & 0.35  & $23.54\pm0.05$ & \citet{2012AA...548A.129F}  \\
  TRGB & $I$ & 0.231 & $23.32\pm0.10$ & \citet{2012AA...540A..49M}  \\
  TRGB & $J$ & 0.23  & $23.31\pm0.05$ & \citet{2011AJ....141..194G} \\
  TRGB & $K$ & 0.23  & $23.26\pm0.07$ & \citet{2011AJ....141..194G} \\
  TRGB & $IJK_s$ & 0.24 & $23.34\pm0.12$ & C\&H (\citeyear{2003ApJ...588L..85C}) \\
\midrule
  RR Lyrae & $V$ & 0.25 & $23.49\pm0.06$ & \citet{2012MNRAS.421.2998F}  \\
  RR Lyrae & $V$ & 0.25 & $23.36\pm0.17$ & \citet{2003ApJ...588L..85C}  \\
\midrule
  Mira & $JHK_s$ & (6)  & $23.56\pm0.03$ & \citet{2013MNRAS.428.2216W} \\
\midrule
  CS & $J$ & 0.35  & 23.52 $\pm$ 0.03 & This work   \\
  CS & $J$ & 0.356 & $23.24\pm0.04$ & \citet{2021ApJ...916...19Z} \\
  CS & $J$ & (7)   & $23.44\pm0.02$ & F\&M (\citeyear{2020ApJ...899...67F}) \\ 
\midrule
\multicolumn{5}{l}{\small{\textit{\textbf{Notes: }}$^*$ statistical errors only, except for C\&H and \citeauthor{2005AA...429..837C}}} \\
\multicolumn{5}{l}{\small{$^{\dag}$Updated in \citet{2021ApJ...916...19Z}}} \\
\multicolumn{5}{l}{\small{$^{\ddag}$Average of multiple studies calibrated with $M_I=-4.05$.}} \\
\multicolumn{5}{l}{\small{(1) optical $+$ $JHK_s$ + mid infrared.}} \\
\multicolumn{5}{l}{\small{(2) $VIJHK_s$ and Infrared Array Camera $3.6\mu m$, and $4.5\mu m$.}} \\
\multicolumn{5}{l}{\small{(3) Not specified.}} \\
\multicolumn{5}{l}{\small{(4) A$_V=0.667$; \; (5) A$_I=0.335$; \; (6) A$_J=0.20$; \; (7) A$_J=0.167$}}\\
\multicolumn{5}{l}{\small{C\&H = \citet{2005AA...429..837C}; \; F\&M = \citet{2020ApJ...899...67F} }}
 \end{tabular}
 \label{tab:dist_n6822}
\end{table}

\begin{figure}
    \centering
    \includegraphics[width=\columnwidth]{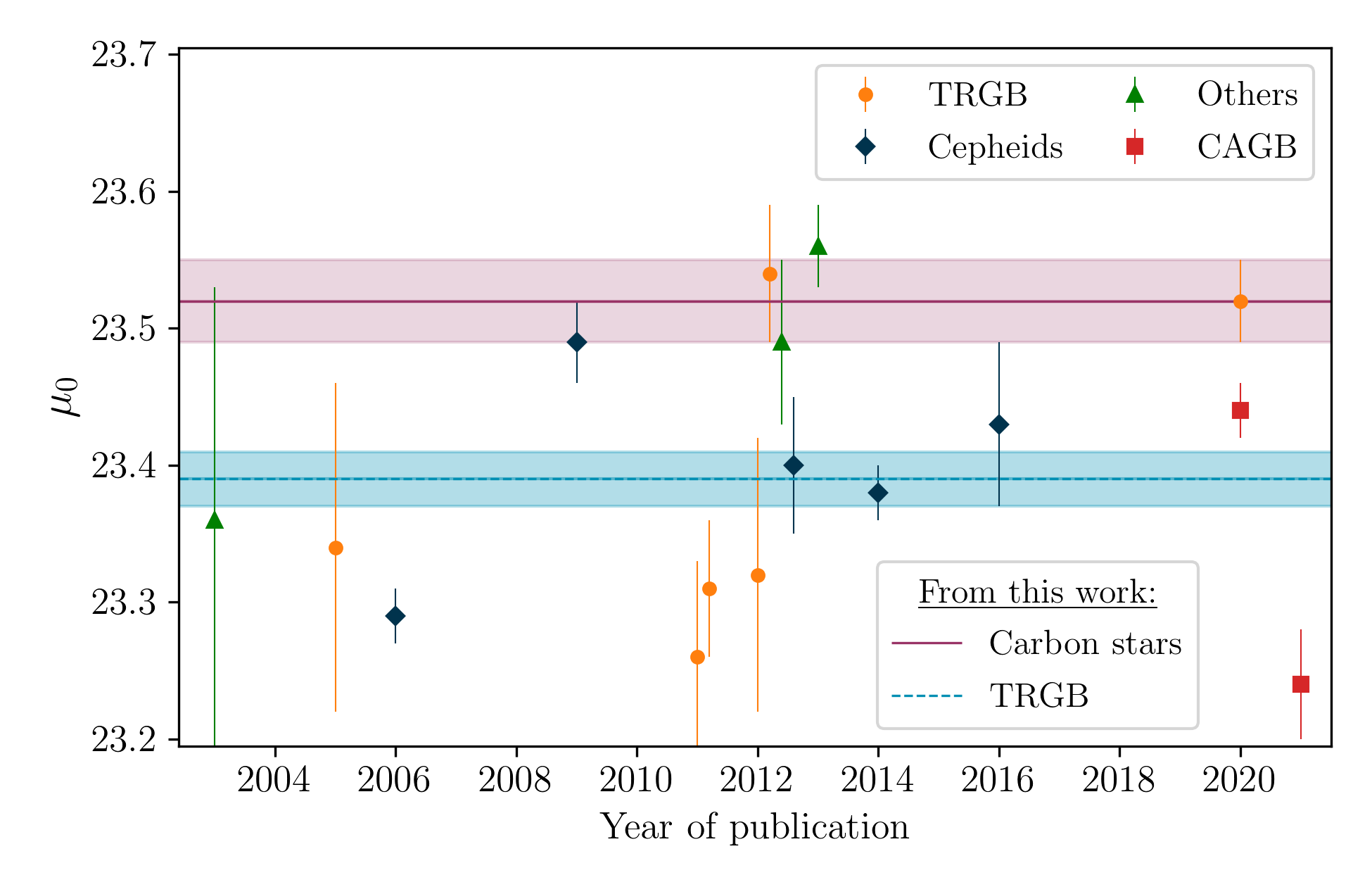}
    \caption{Recent independent distance estimations to NGC 6822 from different techniques. The different methods are plotted using different colours and marker shapes. Others include Miras and RR Lyrae methods. The horizontal lines  covering the full range of years indicate the distance modulus obtained in this work from the CS (straight purple line) and TRGB (dashed blue line) methods. The shaded areas depict the error bars for our results.}
    \label{fig:dis_n6822}
\end{figure}

To compare the different results, the distance moduli listed in Table \ref{tab:dist_n6822} are plotted in Fig. \ref{fig:dis_n6822} against the year of publication. We can see in the plot that, out of the 16 values found in the literature, two of the results agree with both CS and TRGB method, five results agree only with the CS, four only with the TRGB, and four do not agree with either of the estimates in this work.

The TRGB estimate by \citet{2012AA...548A.129F} uses the same extinction correction as in this work and agrees well with the our result obtained using CS. One of the latest TRGB results reported by \cite{2020ApJ...899...67F} is a compilation of $I$-band $\mu_0(TRGB)$ results from the literature. To take an average \citeauthor{2020ApJ...899...67F} standardized the independent estimates using $A_I=0.355$ and $M_I^{TRGB}=-4.05$, giving an average of $\mu_0(TRGB)=23.52$, in good agreement with our CS result.

Returning to the three estimates using CS, we explore the possible sources of the discrepancy between the three values. To calibrate the distances, we use the median $J$ magnitude of the CS in the LMC, $\mathrm{\overline{M}}_{J}=-6.256$. \citet{2021ApJ...916...19Z} instead use the mean $J$ magnitude of the CS in the LMC $M_{J}$=-6.212, while \citet{2020ApJ...899...67F} use the average of the means of the $J$ magnitudes of the CS in the LMC and SMC, $M_{J}$=-6.20. All three calibrations are based on the same distance to the LMC. We find that the mean magnitude of the CS in this study is -6.217 in good agreement with the value found by \citeauthor{2021ApJ...916...19Z}. The mean values for the apparent magnitudes of the CS in NGC 6822 were found to be 17.031 by \citeauthor{2021ApJ...916...19Z} and 17.15 by \citeauthor{2020ApJ...899...67F}. The NGC 6822 CS selected in this work have a mean of 17.32, if we use the mean magnitude our distance estimation would be 23.53, in excellent agreement with the estimate from the median. Another possible source of the discrepancy is in the different colour and magnitude limits used to select the CS. For example \citeauthor{2021ApJ...916...19Z} use a colour selection in the LMC, SMC and WLM of $1.3<(J-K)_0<2.0$ mag while for two other galaxies the blue cut is set at 1.45. For every galaxy in their sample, their selection box has a width $\Delta J_0=2.5$ mag. No colour or magnitude limits are specified for NGC 6822. \citeauthor{2020ApJ...899...67F} employ similar colour constrains but their magnitude range seems significantly narrower than \citeauthor{2021ApJ...916...19Z} and this work. Another possibility are systematic differences between the individual NGC 6822 data sets. For example, the near-infrared data employed by \citeauthor{2021ApJ...916...19Z} are the same data used by \citet{2006ApJ...647.1056G} to determined the distance to NGC 6822 using Cepheids. \citet{2009ApJ...693..936M} found a systematic shift in the zero-point of this $J$ and $K$ photometry and removed these data from their analysis. \citeauthor{2009ApJ...693..936M} state that inclusion of the $J$ and $K$ data in their analysis would have resulted in a 0.17 mag brighter (i.e. a shorter distance) distance modulus. The data set from \citeauthor{2006ApJ...647.1056G}, is also used by \citet{2011AJ....141..194G}, which explains the agreement between the results from these two studies and also with \citeauthor{2021ApJ...916...19Z}.

% -----------------------------------------------------
% -----------------------------------------------------
\subsection{IC 1613}
\label{sec:dis_ic1613}

\begin{figure}
    \centering
    \includegraphics[width=\columnwidth]{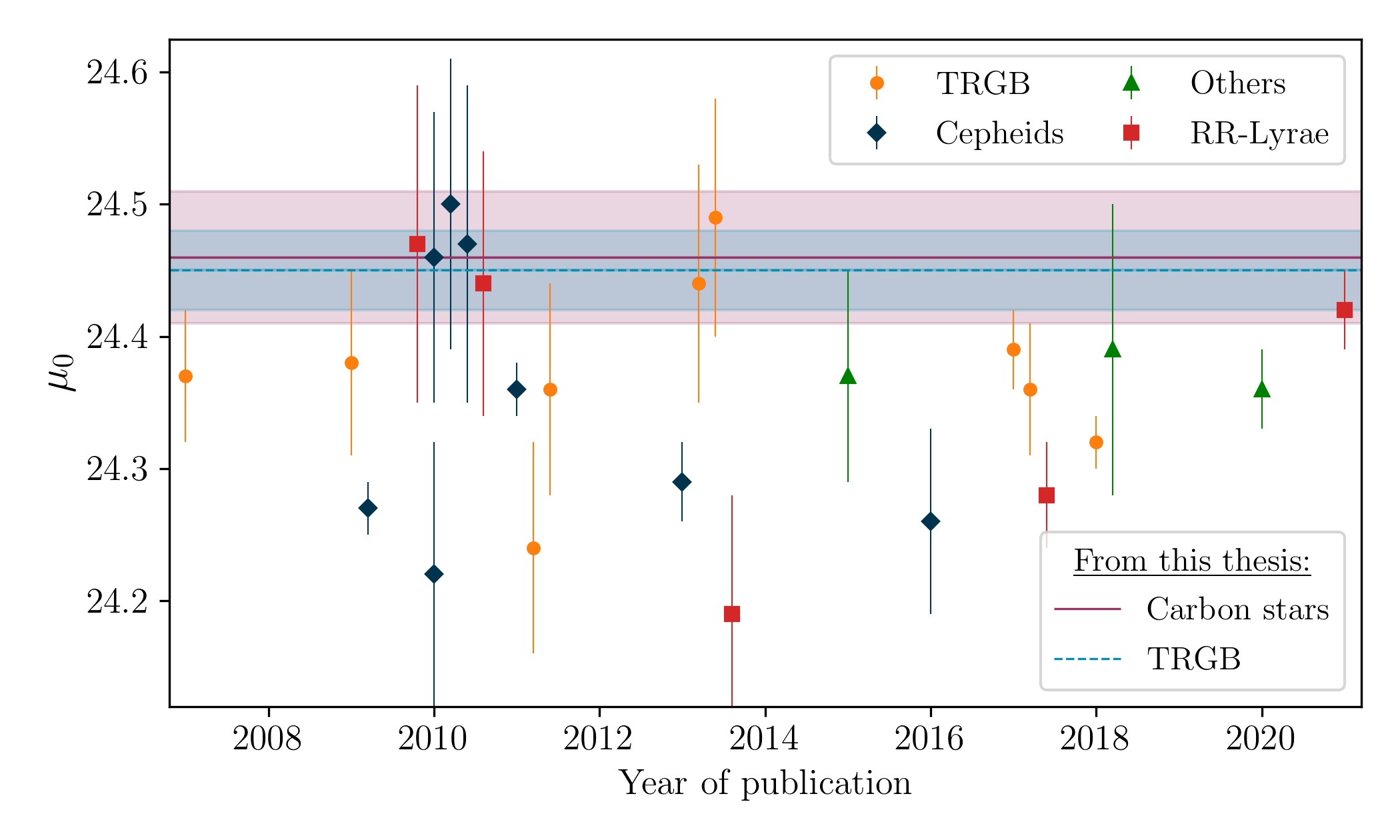}
    \caption{Similar to Fig. \ref{fig:dis_n6822} but for IC 1613. Others include Miras, FGLR and CS methods.}
    \label{fig:dis_ic1613}
\end{figure}

Table \ref{tab:dist_ic1613} presents a summary of recent independent distance estimates for IC 1613. As for Table \ref{tab:dist_n6822}, the values are taken directly from their source without any correction for the use of different extinction values and/or zero-points for the calibration of the distance. The results cover a significant range of values with differences of up to $\sim0.3$ mag. Given the long list of available distances measured to IC 1613 we again plot the values for easier comparison in Fig. \ref{fig:dis_ic1613}. Ten of the 24 individual distance estimates fall outside the range (including error bars) of the distances for IC 1613 from this work. The rest agree within the error bars. No relation between the method used and $\mu_0$ is observed in the plot. From Table \ref{tab:dist_ic1613}, it is also possible to discard any relation between distance measured and the band used to estimate it.

\begin{table}
 \caption[Independent distance estimations to IC 1613]{Independent distance estimations to IC 1613.}
 \centering
 \begin{tabular}{@{\hskip 0pt}c@{\hskip 8pt} c@{\hskip 8pt} c@{\hskip 8pt} c@{\hskip 8pt} c@{\hskip 0pt}}
\midrule 
Method & Band & E$(B-V)$ & $\mu_0^*$ & Reference \\
\midrule 
\midrule 
   Cepheids & near-IR & (3) & $24.26\pm0.07$ & \citet{2016AJ....151...88B}  \\
   Cepheids & (1)   & 0.050 & $24.29\pm0.03$ & \citet{2013ApJ...773..106S}  \\ 
   Cepheids & $BVI$ & 0.025 & $24.34\pm0.03$ & \citet{2011AA...531A.134T}   \\
   Cepheids & $VI$  & (3)   & $24.22\pm0.10$ & \citet{2010ApJ...715..277B}  \\
   Cepheids & $VI$  & 0.025 & $24.47\pm0.12$ & \citet{2010ApJ...712.1259B}  \\
   Cepheids & (2)   & 0.08  & $24.27\pm0.02$ & \citet{2009ApJ...695..996F}  \\
\midrule 
   TRGB & $K_s$ & 0.021 & 24.45$\pm$ 0.02 & This work \\
   TRGB &$JHK$& 0.03  & $24.32\pm0.02$      & \citet{2018ApJ...858...11M}   \\
   TRGB & $I$ &  (4)  & $24.36\pm0.05^{\dag}$& \citet{2017ApJ...845..146H}  \\ 
   TRGB & $I$ &  (5)  & $24.39\pm0.03$      & \citet{2017ApJ...834...78M}   \\
   TRGB & $VI$& 0.021 & $24.44\pm0.09$      & \citet{2013MNRAS.432.3047B}   \\
   TRGB & $VI$& 0.021 & $24.49\pm0.09$      & \citet{2013MNRAS.432.3047B}   \\
   TRGB & $J$ & 0.02  & $24.36\pm0.08$      & \citet{2011AJ....141..194G}   \\
   TRGB & $K$ & 0.02  & $24.24\pm0.08$      & \citet{2011AJ....141..194G}   \\
   TRGB & $I$ & 0.025 & $24.38\pm0.06$      & \citet{2009AJ....138..332J}   \\
   TRGB & $I$ & 0.025 & $24.37\pm0.05$      & \citet{2007ApJ...661..815R}   \\
\midrule 
   RR Lyrae & $VI$ & (3) & $24.42\pm0.03$ & \citet{2021arXiv211106899N}  \\
   RR Lyrae & $V$  & (6) & $24.28\pm0.04$ & \citet{2017ApJ...845..146H}  \\
   RR Lyrae & $V$  & (7) & $24.19\pm0.09$ & \citet{2013MNRAS.435.3206D}  \\
   RR Lyrae & $VI$ &0.025& $24.46\pm0.11$ & \citet{2010ApJ...712.1259B}  \\
\midrule 
   Miras & $JHK_s$ & 0.06 & $24.37\pm0.08$ & \citet{2015MNRAS.452..910M}  \\
\midrule 
   FGLR$^{\ddag}$ & spec. & (7) & $24.39\pm0.11$ & \citet{2018ApJ...860..130B} \\
\midrule 
  CS & $J$ & 0.021 & 24.46 $\pm$ 0.05 & This work \\
  CS & $J$ & (8) & $24.36\pm0.03 $ & F\&M (\citeyear{2020ApJ...899...67F})\\ 
\midrule 
\multicolumn{5}{l}{\small{\textit{\textbf{Notes:}} $^*$ statistical errors only, except for \citeauthor{2010ApJ...712.1259B}}} \\
\multicolumn{5}{l}{\small{$^{\ddag}$flux-weighted gravity-luminosity relationship.}}\\
\multicolumn{5}{l}{\small{$^{\dag}$Updated in F\&M = \citet{2020ApJ...899...67F}}} \\
\multicolumn{5}{l}{\small{(1) $JHK_s$ and Infrared Array Camera $3.6\mu m$, and $4.5\mu m$}} \\
\multicolumn{5}{l}{\small{(2) $BVRIJHK$ and Infrared Array Camera $3.6\mu m$, and $4.5\mu m$}} \\
\multicolumn{5}{l}{\small{(3) Not specified; \; (4) A$_I=0.038$; \; (5) A$_V=0.07$.}}\\
\multicolumn{5}{l}{\small{(6) No corrections applied}}\\
\multicolumn{5}{l}{\small{(7) Estimated individually for each star; \; (8) A$_J=0.018$}}\\
 \end{tabular}
 \label{tab:dist_ic1613}
\end{table}

Recently, \cite{2021arXiv211106899N} used Gaia EDR3 data to estimate distances to several dwarf galaxies based on RR Lyrae variable stars. In their study, \citeauthor{2021arXiv211106899N} report $\mu_0 = 24.42 \pm 0.03$ for IC 1613. This value is in agreement with our results within the statistical error bars. \citeauthor{2021arXiv211106899N} also do an extensive comparison with other distance estimates for IC 1613 published over the last decade. They find that pre-2014 estimates are closer to $\mu \sim 24.45$, compared to $\mu \sim 24.3$ for more recent estimates.

% -----------------------------------------------------
% -----------------------------------------------------
\subsection{WLM}

Table \ref{tab:dist_wlm} presents a compilation of previous distance estimates for WLM based on different standard candles. Again, there is a difference between results prior and after 2014 with distance modulus means of 25.03 and 24.95 respectively. The estimates following 2014 have a smaller dispersion with a standard deviation of 0.02 mag compared to 0.09 mag for the values obtained before 2014. Except for the earliest Cepheid value given, all the values from Table \ref{tab:dist_wlm} fall within the error bars of the estimates from this work. All the estimates published post 2020 are in good agreement with the our results within the error bars, except for \citet{2021ApJ...907..112L}'s TRGB$_I$ distance. For this estimate, \citeauthor{2021ApJ...907..112L} report a systematic error of 0.06 mag, which would account for the small difference. 

\begin{table}
    \centering
    \caption[Independent distance estimations to WLM]{Independent distance estimations to WLM..} 
 \begin{tabular}{@{\hskip 0pt}c@{\hskip 6pt} c@{\hskip 6pt} c@{\hskip 6pt} c@{\hskip 7pt} c@{\hskip 0pt}}
\midrule 
Method & Band & E$(B-V)$ & $\mu_0^*$ & Reference \\
\midrule 
\midrule
Cepheids & $VIJHK$ & 0.05 & 24.98$\pm$0.03 & \citet{2021ApJ...907..112L} \\
Cepheids & near-IR & (1)  & 24.92$\pm$0.07 & \citet{2016AJ....151...88B} \\
Cepheids & $BVI$  & 0.037 & 24.95$\pm$0.03 & \citet{2011AA...531A.134T}  \\
Cepheids & $VI$   & (1)   & 25.09$\pm$0.07 & \citet{2010ApJ...715..277B} \\
Cepheids & $VIJK$ & 0.082 & 24.92$\pm$0.04 & \citet{2008ApJ...683..611G} \\
Cepheids & $VI$   & 0.02  & 25.14$\pm$0.03 & \citet{2007AJ....134..594P} \\
\midrule
TRGB    & $K_s$ & 0.03 & 25.01 $\pm$ 0.03 & This work \\
TRGB & $JHK$ & (2) & 24.98$\pm$0.04 & \citet{2021ApJ...907..112L}  \\
TRGB & $I$ & (3)   & 24.93$\pm$0.02 & \citet{2021ApJ...907..112L}  \\ 
TRGB & $I$ & 0.038 & 24.93$\pm$0.02 & \citet{2019MNRAS.490.5538A}  \\ 
TRGB & $I$ & (4)   & 24.94$\pm$0.03 & \citet{2017ApJ...834...78M}  \\ 
TRGB & $J$ & 0.04  & 25.14$\pm$0.09 & \citet{2011AJ....141..194G}  \\
TRGB & $K$ & 0.04  & 25.12$\pm$0.09 & \citet{2011AJ....141..194G}  \\
TRGB & $I$ & 0.036 & $24.95\pm0.07$ & \citet{2009AJ....138..332J}  \\
TRGB & $I$ & (5)   & $24.93\pm0.04$ & \citet{2007ApJ...661..815R}  \\
\midrule 
FGLR$^{\dag}$ & spec. & (6) & 24.99$\pm$0.10 & \citet{2008ApJ...684..118U}  \\
\midrule
CS & $J$ & 0.03  & 25.03 $\pm$ 0.04 & This work \\
CS & $J$ & 0.082 & 24.95$\pm$0.08 & \citet{2021ApJ...916...19Z} \\
CS & $J$ & (2)   & 24.97$\pm$0.02 & \citet{2021ApJ...907..112L} \\
CS & $J$ & (2)   & 24.97$\pm$0.05 & F\&M (\citeyear{2020ApJ...899...67F}) \\ 
\midrule
\multicolumn{5}{l}{\small{\textit{\textbf{Notes:}} $^*$ statistical errors only.}} \\
\multicolumn{5}{l}{\small{$^{\dag}$flux-weighted gravity-luminosity relationship.}} \\
\multicolumn{5}{l}{\small{(1) Not specified; \; (2) A$_J=0.027$, A$_H=0.017$, A$_K=0.012$}}\\ 
\multicolumn{5}{l}{\small{(3) A$_I=0.057$; \; (4) A$_V=0.1$; \; (5)A$_I=0.07$}} \\
\multicolumn{5}{l}{\small{(6) determined individually for each star.}} \\
\multicolumn{5}{l}{\small{F\&M = \citet{2020ApJ...899...67F} }}
    \end{tabular}
    \label{tab:dist_wlm}
\end{table}

% -----------------------------------------------------
% -----------------------------------------------------

\subsection{NGC 3109}

Independent distance moduli determinations for NGC 3109 are presented in Table \ref{tab:dist_n3109}. Of the fourteen estimates found in the recent literature, ten are in very good agreement with with our results.

\begin{table}
    \centering
    \caption{Independent distance estimations to NGC 3109.} 
 \begin{tabular}{@{\hskip 0pt} c@{\hskip 8pt} c@{\hskip 8pt} c@{\hskip 8pt} c@{\hskip 8pt} c@{\hskip 0pt} }
\midrule 
Method & Band & E$(B-V)$ & $\mu_0^*$ & Reference \\
\midrule 
\midrule
Cepheids & near-IR & (1)& 25.49$\pm$0.06 & \citet{2016AJ....151...88B} \\
Cepheids & $VI$ & (1)   & 25.58$\pm$0.12 & \citet{2010ApJ...715..277B} \\
Cepheids & $JK$ & 0.087 & 25.57$\pm$0.02 & \citet{2006ApJ...648..375S} \\ 
Cepheids & $VI$ & 0.05  & 25.54$\pm$0.03 & \citet{2006ApJ...648..366P} \\
\midrule
TRGB    & $K_s$ & 0.056 & 25.58 $\pm$ 0.03 & This work\\
TRGB & $J$  & 0.06  & 25.49$\pm$0.05 & \citet{2011AJ....141..194G} \\ 
TRGB & $K$  & 0.06  & 25.42$\pm$0.05 & \citet{2011AJ....141..194G} \\ 
TRGB & $V$  & (2)   & 25.55$\pm$0.05 & \citet{2009ApJS..183...67D} \\ 
TRGB & $I$  & (2)   & 25.50$\pm$0.04 & \citet{2009ApJS..183...67D} \\ 
TRGB & $I$  & 0.058 & 25.62$\pm$0.09 & \citet{2009AJ....138..332J} \\ 
TRGB & $VI$ & (3)   & 25.61$\pm$0.10 & \citet{2008AJ....136.2332H} \\
TRGB & $I$  & (4)   & 25.56$\pm$0.05 & \citet{2007ApJ...661..815R} \\
\midrule
FGLR$^{\dag}$ & spec. & (5) & 25.55$\pm$0.09 & \citet{2014ApJ...785..151H} \\ 
\midrule 
CS      & $J$   & 0.056 & 25.59 $\pm$ 0.03 & This work \\
CS & $J$ & 0.087 & 25.52$\pm$0.05 & \citet{2021ApJ...916...19Z} \\
CS & $J$ & (6)   & 25.56$\pm$0.05 & F\&M (\citeyear{2020ApJ...899...67F}) \\ 
\midrule
\multicolumn{5}{l}{\small{\textit{\textbf{Notes:}} $^*$ statistical errors only.}} \\
\multicolumn{5}{l}{\small{$^{\dag}$flux-weighted gravity-luminosity relationship.}} \\
\multicolumn{5}{l}{\small{(1) Not specified; \; (2) A$_V=0.201$; (3) A$_V=0.221$, A$_I=0.129$}} \\
\multicolumn{5}{l}{\small{(4) A$_I=0.13$ \; (5) determined individually for each star.}}\\ 
\multicolumn{5}{l}{\small{(6) A$_J=0.027$; \; F\&M = \citet{2020ApJ...899...67F} }} \\
    \end{tabular}
    \label{tab:dist_n3109}
\end{table}

%%%%%%%%%%%%%%%%%%%%%%%%%%%%%%%%%%%%%%%%%%%%%%%%%%%%%%%%%%%%%%%%
%%%%%%%%%%%%%%%%%%%%%%%%%%%%%%%%%%%%%%%%%%%%%%%%%%%%%%%%%%%%%%%%

\section{Discussion}

\newerased{Carbon stars (CS) are a potentially powerful distance indicator.} In Paper II we developed a method to estimate distances to Magellanic type galaxies using the median $J$ magnitude of CS and tested it in two galaxies. In this paper we have refined our method and expanded our sample of galaxies to four. \newerased{We also compared the results from the CS with an estimate of the distance using the TRGB edge-detection method.} To determine the distance modulus we relied on two different calibrators, the LMC and SMC. To obtain the absolute magnitudes of the CS for the Magellanic Clouds we adopted the mean distance modulus  $\mu_{\mathrm{LMC}}$ = 18.477 $\pm$ 0.004 (statistical) $\pm$ 0.026 (systematic) \citep{2019Natur.567..200P} for the LMC, and $\mu_{\mathrm{SMC}}$ = 18.977 $\pm$ 0.016 (statistical) $\pm$ 0.028 (systematic) \citep{2020ApJ...904...13G} for the SMC. 

\newinserted{In the current paper we have changed the reddening maps for the Magellanic Clouds from \citet{2020ApJ...889..179G} to those more recently published by \citet{2021ApJS..252...23S}. If we compare with Paper I, correcting the data using the \citeauthor{2021ApJS..252...23S} reddening maps produces fainter magnitudes for the LMC stars, with the most noticeable effect on the TRGB $K_s$ magnitude. If we repeat the TRGB analysis on the LMC using the \citeauthor{2020ApJ...889..179G} maps the TRGB magnitude in the $K_s$ band is $\sim$~0.1 magnitudes brighter than the value obtained using the \citeauthor{2021ApJS..252...23S} reddening maps. No significant effect is seen for the SMC. The extinction coefficients ($R_x$) have also been updated to the values given in \citet{2022MNRAS.511.1317C}. The main difference between the coefficients used in Paper I from \citet{2003ApJ...594..279G} and \citeauthor{2022MNRAS.511.1317C} is the value of $R$ for the $K$ filter which is $\sim$4 times larger in the newer study. The difference in $R_K$ leads to a significant shift in the $(J-K_s)_0$ colour of the stars with the red giants moving approximately $\sim$0.05 magnitudes towards the red.} In the case of the CS, using the older extinction coefficients from \citeauthor{2003ApJ...594..279G} produced fainter median $J$ magnitudes compared with those obtained using \citeauthor{2022MNRAS.511.1317C}, with a difference of 0.010 mag for the LMC and 0.021 mag for the SMC. By contrast, the TRGB $K_s$ magnitude seems less affected by the use of different extinction coefficients with the old coefficients yielding values of 0.004 and 0.010 magnitudes fainter for the LMC and SMC respectively. The difference between the values obtained with the newer \citeauthor{2022MNRAS.511.1317C} and older \citeauthor{2003ApJ...594..279G} coefficients translate into slightly larger distance moduli: galaxies that use the SMC CS as calibrator have  the distance moduli larger by 0.021 magnitudes, and those that use the LMC are larger by 0.010 magnitudes. For both the CS and the TRGB methods, when using different extinction coefficients, the distance modulus still falls within the error bars of the corresponding values.

Our approach in establishing the magnitude limits over which the CS luminosity function extends is generally to make the most benign selection possible. Since the CS are redder than most other AGB stars, and since the Magellanic Cloud CS are generally brighter than background galaxies, no magnitude cut is required to isolate the sample in these two galaxies. This can be seen quite clearly in Fig \ref{fig:CSselMCs}. If there is any intrinsic asymmetry in the CS LFs from galaxy to galaxy, this selection process will not suppress it. In the case of the target galaxies examined here, they are distant enough that faint background galaxies are a potential source of contamination. In these cases only a mild faint magnitude cut was made and we showed in Paper II that this did not affect the distance determination. Additionally, to preserve as much information as possible in the luminosity function, we used the median $J$ magnitude as our standard candle.

For the CS we have shown in Papers I and II that the $J$ luminosity functions of the LMC and SMC have different shapes. The differences were attributed to differences in the metallicities and star formation histories of the two galaxies. The main observable difference is that the LMC CS luminosity function is skewed towards brighter magnitudes. To quantify this skewness we modelled the CS distributions using a modified Lorentzian model that allows for skewness as well as different weights on the tails of the distributions. The results for the skew parameters confirmed that the LMC ($s\sim-0.4$) and SMC ($s\sim0.0$) have different shapes, with the LMC skewed towards brighter magnitudes. The LMC also shows a more extended tail compare to the SMC, while the widths of the distributions are almost identical. The fitting of the model was made directly to the binned luminosity function using non-linear least squares. Given the large number of stars in the Magellanic Clouds, this fitting method gave reasonable-sized error bars for the parameters that define the model. This was not the case for the target galaxies as the number of CS available is considerably smaller. To improve the error bars in the models for our sample of galaxies in this paper, we developed an un-binned maximum likelihood method (MLE) to fit the modified Lorentzian model. The MLE methods reduced the error bars in the parameters of the Lorentzian model by at least a few percent, but for the skew parameter the error bars improved by an average of $\sim75\%$. The parameters obtained from the linear squares fit and the MLE method are consistent within the error bars.

\newerased{ To classify the target galaxies as "LMC-like" or "SMC-like" we used the skew parameter. Following the condition defined in equation \ref{eq:dm}, NGC 6822 was classified as an "LMC-like" galaxy, while IC 1613, WLM and NGC 3109 were classified as "SMC-like". The distance moduli obtained for these galaxies using the most suitable calibrator are provided in Table \ref{tab:dm_results}. }

\newinserted{From our analysis we can estimate the contribution of several possible sources of systematic errors. The first is the fact that different galaxies may have different distributions of carbon stars. This was the focus of Paper II. We have attempted to address this particular error, by using the LMC and SMC to estimate the size of this error. In Paper II we found that the systematic uncertainty introduced by using the two different calibrators was only $\sim$0.1 mag. With the new reddening maps this error was reduced to 0.07 in distance modulus for the CS technique.}

\newinserted{Looking at other galaxies in our sample and comparing techniques (Table~\ref{tab:dm_results}), if we restrict the analysis to ``SMC-like'' galaxies (as determined from the CS luminosity function), the agreement between techniques is striking at about 0.02 in distance modulus. However, for the ``LMC-like'' galaxy NGC 6822, the two techniques differ by 0.13 indicating possible systematic errors with ``LMC-like'' (perhaps higher metallicity) galaxies and with the treatment of reddening. On the other hand, the SMC-like galaxies in the sample indicate that when the same data set is used with the same reddening and colour corrections, the estimates of the distance modulus agree within the statistical errors.   This indicates that among galaxies that we can identify as ``SMC-like'' there is a striking uniformity in the luminosity function of the giant stars from galaxy to galaxy. Furthermore, the potential systematic errors introduced by correction for absorption and band corrections can be mitigated by looking at regions of the sky with small Galactic reddening and using the same filter set (and hopefully instrument) for the local calibration and the observations of distant galaxies. Although we find striking agreement between our two measurement techniques among the ``SMC-like'' galaxies, Table~\ref{tab:dist_n6822} through~\ref{tab:dist_n3109} indicate that different studies using different techniques typically find distance moduli for the same galaxies that differ from study to study by much more than the quoted statistical errors. Taking these results at face value, one would estimate that the systematic errors are about twice the quoted statistical errors for a reddened ``LMC-like'' galaxy  like NGC~6822 and comparable to the statistical errors for SMC-like galaxies.}

The CS method is a distance determination technique still in development. Nevertheless, further standardization and calibration of the CS method might be able to compensate for the difference in the skewness of the CS luminosity functions and allow for a global calibration. Most standard candles in use today require some standardization and calibration. For example, SN Ia exhibit light-curves with maximum intrinsic luminosities that differ by more than 1 magnitude \citet{1993ApJ...413L.105P, 1995AJ....109....1H}. SN Ia distances are then calibrated with techniques that use the multicolour light-curve shapes \citet{1996ApJ...473...88R} and the stretch method \citet{1997ApJ...483..565P}.

\newerased{ We also estimated the distances to our galaxies using the TRGB edge-detection method. Because of the absence of a parametrized model, we could not establish a method to quantify the difference in the shape of the red-giant luminosity function, instead we relied on the classification obtained from the CS luminosity function. The distance moduli estimated using the TRGB edge-detection method are provided in Table \ref{tab:dm_results}.}

\newerased{Except for the result obtained from the TRGB for NGC 6822, the distance moduli obtained from the CS median $J$ magnitude and TRGB-$K_s$ magnitude agree well within the statistical error bars.}

\newinserted{In our analysis of the TRGB we observed that }
the response function for the LMC shows a second peak similar in size to the first peak which is detected as the TRGB in $\sim30\%$ of the bootstrap iterations. This second peak is 0.06 magnitudes brighter than the first peak. \cite{2019ApJ...880...63M} used simulated photometry to investigate the dependency of the TRGB luminosity on stellar age and metallicity as a function of wavelength. In the $J$ and $K_s$-band, using a transformed magnitude similar to the $T$ magnitude, \citeauthor{2019ApJ...880...63M} find that a $\Delta m_{K_s}\sim0.04$ magnitude difference remains after correcting for metallicity effects. The difference found theoretically is only slightly smaller than what we find between SMC and the second peak detected in the LMC. 

Since our distance determination method relies on an ``LMC/SMC-like'' classification, it has thus only been tested on Magellanic type galaxies. An analysis of the CS in the Milky Way carried out in Paper I showed that the median $J$ magnitude was considerable fainter (-5.60) compared to the Magellanic Clouds. Recently, \cite{cat_lee} derived a median $J$ magnitude of -6.14 based on Gaia EDR3. The two values are not in agreement but it is necessary to consider that Paper I and \citeauthor{cat_lee} used two different CS catalogues with \citeauthor{cat_lee} using a much smaller and more highly selected sample. Such a selection might not be available in distant galaxies where the ultimate plan is to use this method for a determination of the Hubble Constant. Part of the difference in the median values found for the Milky Way could be attributed to the fact that Paper I  calibrates the absolute magnitudes using Gaia DR2.  Further analysis on a larger sample of Milky Way CS will be carried out in a future paper. 

%%%%%%%%%%%%%%%%%%%%%%%%%%%%%%%%%%%%%%%%%%%%%%%%%%%%%%%%%%%%%%%%
%%%%%%%%%%%%%%%%%%%%%%%%%%%%%%%%%%%%%%%%%%%%%%%%%%%%%%%%%%%%%%%%
%   _____                ___          _
%  / ____|               | |         (_)                
% | |     ___  _ __   ___| |_   _ ___ _  ___  _ __  ___ 
% | |    / _ \| '_ \ / __| | | | / __| |/ _ \| '_ \/ __|
% | |___| (_) | | | | (__| | |_| \__ \ | (_) | | | \__ \
%  \_____\___/|_| |_|\___|_|\__,_|___/_|\___/|_| |_|___/
%                                                       
%                                                
\section{Conclusions}

In this work, we have refined our independent distance determination method developed in Paper II. Our method uses the median $J$ magnitude of the carbon-rich AGB star and relies on the determination of the most suitable calibrator (either the LMC or SMC) to obtain the distance modulus. We also compared the results from the CS with an estimate of the distance using the TRGB edge-detection method.

In this paper the method was tested in four members of the Local Group: NGC 6822, IC 1613, WLM and NGC 3109. All galaxies were corrected for extinction before selecting the CS or RGs. This is especially important for the CS as these stars are selected based solely on their $(J-K_s)_{0}$ colour. The use of extinction corrected colours allows selecting the CS using the same colour limits for all the target galaxies.

The most suitable calibrator for each target galaxy is chosen based on the characteristics of the luminosity function of the CS. 
To characterise the luminosity functions of the CS we model each function using a modified Lorentzian model fitted using an un-binned maximum likelihood method. The modification to the Lorentzian allows the model to account for asymmetry outputting a skew parameter value which we use to determine whether a target galaxy is ``LMC-'' or ``SMC-like''. The determination of the most suitable calibrator is important as choosing the wrong Magellanic Cloud as calibrator introduces a systematic error on the distance
modulus of $\sim$0.1 mag.

Following the condition defined in equation \ref{eq:dm}, NGC 6822 was classified as an "LMC-like" galaxy, while IC 1613, WLM and NGC 3109 were classified as "SMC-like". The distance moduli obtained for these galaxies using the most suitable calibrator are provided in Table \ref{tab:dm_results}. 

We also estimated the distances to our galaxies using the TRGB edge-detection method. Because of the absence of a parametrized model, we could not establish a method to quantify the difference in the shape of the red-giant luminosity function, instead we relied on the classification obtained from the CS luminosity function. The distance moduli estimated using the TRGB edge-detection method are also provided in Table 9.

Except for the result obtained from the TRGB for NGC 6822,
the distance moduli obtained from the CS median $J$ magnitude and
TRGB-$K_s$ magnitude agree well within the statistical error bars.

%%%%%%%%%%%%%%%%%%%%%%%%%%%%%%%%%%%%%%%%%%%%%%%%%%%%%%%%%%%%%%%%
%%%%%%%%%%%%%%%%%%%%%%%%%%%%%%%%%%%%%%%%%%%%%%%%%%%%%%%%%%%%%%%%
\section*{Acknowledgements}

This work was supported by the Natural Sciences and Engineering Research Council of Canada, the Canada Foundation for Innovation, the British Columbia Knowledge Development Fund. Part of this research is based on observations obtained with WIRCam, a joint project of CFHT, Taiwan, Korea, Canada, France, at the Canada-France-Hawaii Telescope (CFHT) which is operated by the National Research Council (NRC) of Canada, the Institut National des Sciences de l'Univers of the Centre National de la Recherche Scientifique of France, and the University of Hawaii. This research has also made use of the services of the ESO Science Archive Facility.

\section*{Data availability}
The LMC and SMC catalogues developed in Paper I can be found at: \url{https://gitlab.com/pripoche/using-carbon-stars-as-standard-candles}. The photometric catalogue for NGC 6822, IC 1613 (WIRCam), WLM and NGC 3109 will be shared upon request to the corresponding author.



\bibliographystyle{mnras}
\bibliography{biblio} % if your bibtex file is called example.bib

%%%%%%%%%%%%%%%%%%%%%%%%%%%%%%%%%%%%%%%%%%%%%%%%%%%%%%%%%%%%%%%%
%%%%%%%%%%%%%%%%%%%%%%%%%%%%%%%%%%%%%%%%%%%%%%%%%%%%%%%%%%%%%%%%
%%%%%%%%%%%%%%%%%%%%%%%%%%%%%%%%%%%%%%%%%%%%%%%%%%%%%%%%%%%%%%%%
%%%%%%%%%%%%%%%%%%%%%%%%%%%%%%%%%%%%%%%%%%%%%%%%%%%%%%%%%%%%%%%%

% Don't change these lines
\bsp	% typesetting comment
\label{lastpage}
\end{document}